\documentclass[12pt]{iopart}
\usepackage{iopams}
\usepackage{setstack}
\usepackage{lipsum}
\usepackage{comment}
\usepackage[hyphens]{url}
\usepackage[hidelinks]{hyperref}
\hypersetup{breaklinks=true}
\usepackage{cite}
\usepackage{siunitx}
\DeclareSIUnit\torr{Torr}
\DeclareSIUnit\in{in}
\usepackage{graphicx}
\usepackage{booktabs}
\usepackage{pifont} %símbolo x del campo magnético
\usepackage{newclude} %include sin pagebreak

\begin{document}
%https://apastyle.apa.org/style-grammar-guidelines/research-publication/dissertation-thesis 

\title[Characterizing SCR-1 Stellarator Physics]{Characterization of the SCR-1 Stellarator Physics: Investigating Plasma Discharge, MHD Equilibrium Calculations, and O-X-B Mode Conversion Feasibility}
%\author{P J Smith$^1$, T M Collins$^2$, R J Jones$^3$\footnote{Present address: Department of Physics, University of Bristol, Tyndalls Park Road, Bristol BS8 1TS, UK.} and Janet Williams$^3$}
%\author{R Solano-Piedra$^{1,2}$, V I Vargas$^{1,2}$, L A Araya-Solano$^{1,2}$, A Köhn-Seemann$^{3}$, F Vílchez-Coto$^{2}$}
\author{R Solano-Piedra$^{1,2}$, V I Vargas$^{1,2}$, L A Araya-Solano$^{1,2}$, F Vílchez-Coto$^{2}$}

\address{$^1$ School of Physics, Costa Rica Institute of Technology, Campus Cartago, CR}
\address{$^2$ Plasma Laboratory for Fusion Energy and Applications, Costa Rica Institute of Technology, Campus Cartago, CR}
%\address{$^3$ Institute of Interfacial Process Engineering and Plasma Technology IGVP, University of Stuttgart, 70569 Stuttgart, Germany}

\ead{risolano@itcr.ac.cr}
\begin{abstract}

The {\it Stellarator de Costa Rica 1} (SCR-1) is a modular stellarator with a small aspect ratio that serves as a valuable research and training tool for plasma magnetic confinement. This study explored a new heating mechanism and the factors that influence it. The current state of the device and plasma discharge are initially presented. Subsequently, the measurement process was utilized to determine radial profiles of electron density and electron temperature using a single Langmuir probe, and the results were compared with theoretical predictions based on the particle and energy balance. Additionally, the VMEC code was employed to calculate magnetic flux surfaces with characteristics such as a low aspect ratio, low beta parameter, negative magnetic shear, and decreasing rotational transform along magnetic flux surfaces. The Mercier criterion was employed to conduct a linear stability analysis, which identified a magnetic well that played a crucial role in the linear stability of the majority of magnetic flux surfaces. Feasibility studies of electron Bernstein waves were conducted using the IPF-FMDC full-wave code, with input files generated from the device and known plasma characteristics. The results obtained from the IPF-FMDC full-wave code revealed that the O-X conversion percentage reached a maximum of 63 \% when considering radiation reflection in the vacuum vessel. Significant effects of plasma curvature on the O-X wave conversion and normalized electron density scale length were observed, while the change in the SCR-1 heating position did not produce a significant impact. Three damping mechanisms affecting O-X conversion were studied, and one of the principal effects was the SX-FX conversion due to steep electron density gradient. Additionally, stochastic electron heating showed a low electron field amplitude, which is important for limiting the electron Bernstein wave propagation.

%\lipsum[1-2]
\end{abstract}
%\vspace{-0.5cm}
%\keywords{magnetic moment, solar neutrinos, astrophysics}
\noindent{\submitto{\PPCF}}
\maketitle
%\ioptwocol %doble columna

\section{Introduction}

Small-scale magnetic confinement devices play a crucial role in advancing towards controlled thermonuclear fusion.  These devices provide a platform for training individuals in various skill areas of plasma fusion research, such as operation, maintenance, and performance optimization. In addition, they are employed as cost-effective testing grounds for exploring different plasma phenomena and heating and diagnostic mechanisms, which can subsequently be scaled up to larger machines to streamline the experimental process \cite{iaea2016}.  

Given these considerations, the SCR-1 stellarator, a small-aspect-ratio modular stellarator, first of its type in Latin America \cite{Vargas2017}, has a research program dedicated to identifying challenges related to both physics and engineering in small-scale magnetic confinement devices. The Plasma Laboratory for Fusion Energy and Applications aims to function as a cost-effective facility for the design and construction of stellarators and as a training center for researchers. This includes the characterization of plasma parameters using a diverse set of plasma diagnostics, optimization of the magnetohydrodynamic (MHD) equilibrium, and enhancements in plasma heating mechanisms. However, the pursuit of these objectives is not without challenges, some of which have been identified in the context of the SCR-1 stellarator, particularly regarding its size, during experimental campaigns and theoretical research. Hence, presenting the current state of the SCR-1 stellarator is essential as it provides comprehensive insights into its peripheral systems and technical plasma discharge processes. This approach aligns with the methodology outlined by \cite{obiki2001, krause2002}.
%These challenges include attaining plasma scaling conditions, designing effective plasma diagnostics, and optimizing the plasma and magnetic field parameters.

Furthermore, the achievement of these goals depends significantly on the administration, proficiency, and development of computational codes relevant to stellarator physics. Attaining  MHD equilibrium and linear stability under the current conditions of the SCR-1 stellarator is a critical and ongoing endeavor to optimize plasma processes and push the limits of the device. The VMEC code \cite{hirshman1983} has emerged as the initial step for computing the MHD equilibrium of nested magnetic flux surfaces as shown in the W7-X stellarator \cite{suzuki2006}. 
%This code has been seamlessly integrated into web services for the W7-X stellarator, enhancing accessibility and facilitating integration with other codes pertinent to plasma characterization \cite{suzuki2006, grahl2018}.

In addition, this computational approach offers opportunities for enhancing the plasma-heating mechanism. One possible approach involves creating an overdense plasma by generating electron Bernstein waves (EBW). This method can be used alongside electron cyclotron resonant heating (ECRH), a commonly used technique for increasing electron temperature and electron density in plasmas \cite{xu2016}. However, ECRH has a limitation where it cannot reach the core of the plasma due to a cutoff density \cite{laqua2007}. The implementation of EBW in a large number of stellarators and tokamaks worldwide, encompassing a diverse spectrum of plasma parameters, has been ongoing since the 2000s to the present day.  This sustained effort has yielded valuable simulations and experimental validations \cite{bilato2009, elserafy2019, freethy2023}.
%One such avenue involves investigating overdense plasma through the generation of electron Bernstein waves (EBW), which can serve as a complementary mechanism to electron cyclotron resonant heating (ECRH) a commonly used method for increasing the electron temperature \cite{xu2016} that comes with the drawback of a cutoff density that does not reach the core of the plasma \cite{laqua2007}

The O-X-B mechanism for EBW research has been a primary focus in devices with plasma length scales larger than the incident radiation wavelength. However, the SCR-1 stellarator did not satisfy these conditions. Therefore, the use of a full-wave code in the simulations is necessary to provide a comprehensive explanation of how mode conversion is used to heat the plasma. This approach facilitates the visualization of wave propagation and possible absorption, whether through resonant or non resonant mechanisms, as well as the computation of O-X mode conversion efficiency. Understanding the factors that determine the maximum O-X conversion efficiency requires a comparison with the relevant theoretical aspects, as discussed by \cite{khusainov2018, shalashov2018}.

This work presents a comprehensive study of the plasma discharge aspects, MHD equilibrium calculations, and feasibility simulation process of the O-X-B mechanism for electron Bernstein wave generation in the SCR-1 stellarator. This paper is organized into four main sections. Section 2 provides an examination of the peripheral systems and plasma discharge dynamics of the SCR-1 stellarator. In Section 3, measurements of electron temperature and electron density are conducted using a simple Langmuir probe, incorporating corrections based on tip dimensions with orders of magnitude similar to the Larmor radius. These measurements yielded electron temperature and density radial profiles that were systematically compared with the theoretical model proposed by \cite{Lechte2002} to validate the acquired data. The magnetohydrodynamic equilibrium and linear stability based on the Mercier criterion in SCR-1 were thoroughly examined in Section 3. Finally, in Section 4, the generation of electron Bernstein waves within the SCR-1 plasma through the O-X-B mechanism is discussed. The results obtained cover a series of microwave heating scenarios involving electronic density and magnetic field modulus variations as well as considerations of the SCR-1 vacuum vessel, delving into greater detail than those shown in \cite{Coto2020}.

\section{The SCR-1 stellarator} \label{sec:SCR1}
\begin{figure}[h!]
    \centering
    \includegraphics[width=0.6\linewidth]{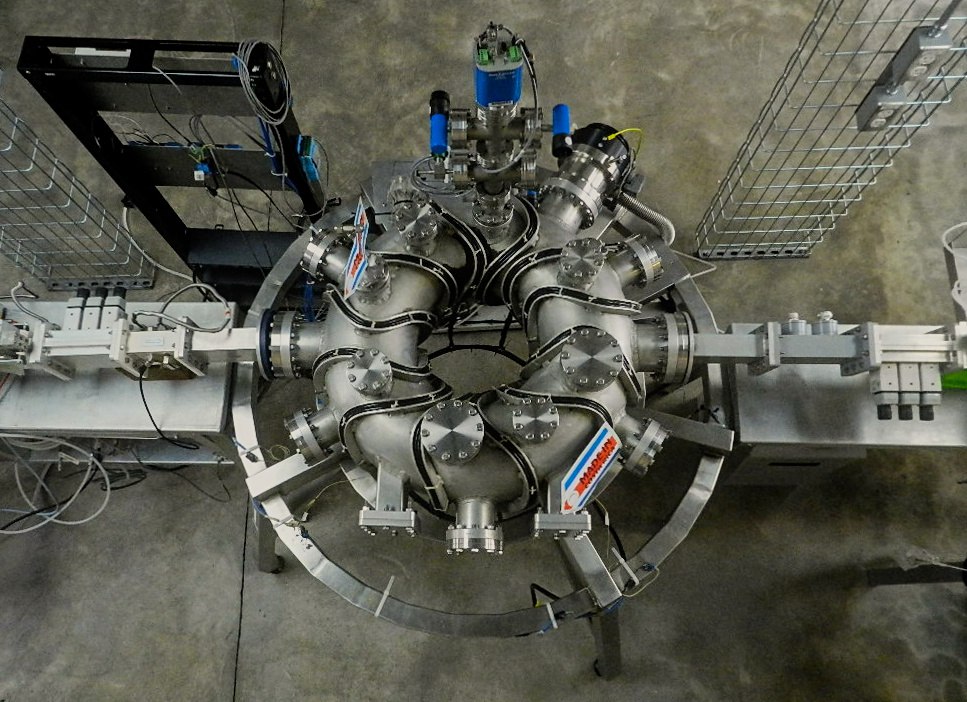}
    \caption{Top view of SCR-1 stellarator with its ECR heating system connections and control and data acquisition system}
    \label{fig:SCR-1}
\end{figure}

The SCR-1, shown in Figure \ref{fig:SCR-1}, field period \num{2}, represents a pioneering achievement in Latin America, serving as the region's inaugural magnetic confinement device-type stellarator for high electron temperature plasmas. A historic milestone was reached on June 29, 2016, with the first public plasma discharge, highlighting a moment of great significance in the realm of applied physics in Costa Rica, with active participation from government entities \cite{hoyeneltec2016}. The SCR-1 stellarator comprises six pivotal systems, each of which is described as follows: vacuum, power supply, gas injection, Electron Cyclotron Resonance Heating, control and data acquisition and Plasma diagnostics.

\subsection{Peripherical systems}\label{{subsec:comp_sonda}}
The vacuum system was composed of a vacuum vessel fabricated from two metal plates that were shaped and bonded via MIG welding. The chosen material is a specialized alloy, denoted as 6061-T6, which is a composite of aluminum, silicon, and magnesium. This vessel is equipped with a total of twenty-four Conflat ports with diameters of \SI{6.000}{\in}, \SI{4.500}{\in}, and \SI{3.375}{\in}, alongside two rectangular ports.  The essential geometric characteristics of the vacuum vessel are presented in Table \ref{table:characteristics_SCR1}. To maintain an operational pressure of approximately \SI{4e-5}{\torr}, the vacuum system was equipped with a high-vacuum ion-gauge manometer, medium- and low-vacuum convectrons, vent valves,  turbomolecular vacuum pump, and mechanical pump, all of which collectively monitored and regulated the requisite conditions.
\begin{table}[h!]
\caption{\label{table:characteristics_SCR1}Technical characteristics of the SCR-1 vacuum vessel}
\begin{indented}
%\centering
\item[] \begin{tabular}{@{}cc}
\br    
Parameter & Magnitude \\
\mr
Thickness (\si{\mm}) & \num{4.0} \\
Volume (\si{\meter^3}) & \num{0.0434} \\
External radius (\si{\mm}) & \num{364.1} \\
Internal radius (\si{\mm}) & \num{112.1} \\
Major radius (\si{\mm}) & \num{247.7} \\
\br
\end{tabular}
\end{indented}
\end{table}

\begin{figure}[h!]
    \centering
    \includegraphics[width=0.4\linewidth]{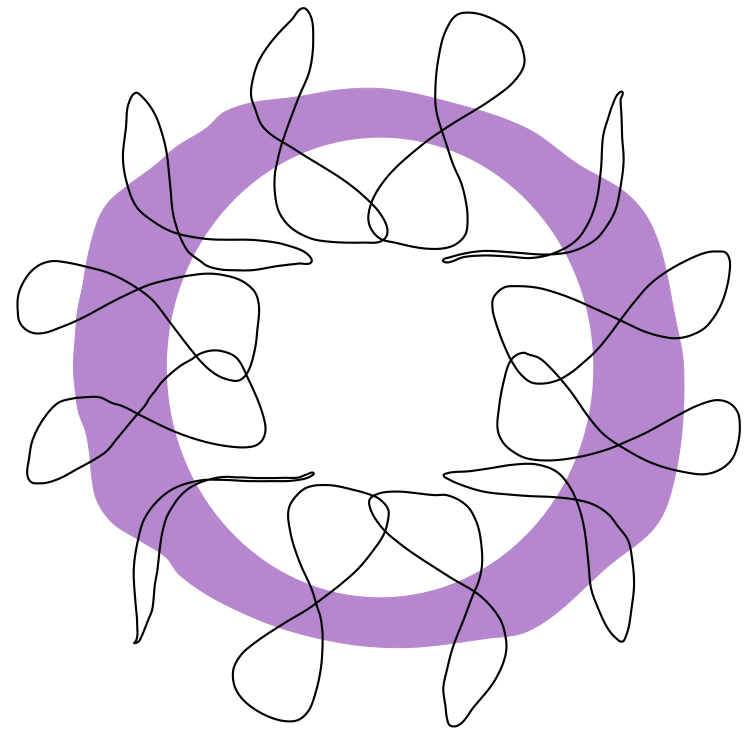}
    \caption{Top view of the last magnetic flux surface with its configuration of coils placed in the toroidal direction.}
    \label{fig:bobinas}
\end{figure}

The power supply system was equipped with a set of \num{12} modular coils, each composed of six turns of AWG\#4 wire per coil. The shape of each coil on the last closed magnetic-flux surface is shown in Figure \ref{fig:bobinas}. These coils had a current of \SI{4350}{\ampere} and were powered by a configuration of 60 batteries, each with a nominal voltage of \SI{2.0}{\volt}, and a battery capacity system rated at \SI{150}{\ampere. \hour} \cite{mora2015}.

The working gas, hydrogen, in the gas injection system was conveyed directly to a mass flow controller that guided it through a single path to the vacuum vessel, which was equipped with needle valves to ensure process maintenance and safety, and a vent valve was necessary to purge the system in each plasma discharge. The mass flow controller was set to provide constant gas flow injection without self-regulation capability.

The electron cyclotron resonant heating system consists of two magnetrons: \SI{2.0}{\kilo\watt} and \SI{3.0}{\kilo\watt}. Plasma heating at \SI{2.45}{\giga\hertz}, with a cutoff electron density ($n_{ecut}$) of \SI{7.15e16}{\metre^{-3}} by the second-harmonic magnetic field ($B_{ce}$) of \SI{43.8}{\milli\tesla}. Continuous emission of electromagnetic waves in the $\text{TE}_{10}$ mode was sustained for an approximate duration of \SI{5}{\second}. A microwave insulator is used to protect the magnetron. During the plasma discharge, a directional coupler was employed to measure both the reflected and transmitted power. This measurement signal then proceeds to a section of the system equipped with a three-stub tuner. In addition, the system includes a waveguide with a quartz window. The quartz window was designed to have a surface flatness equal to half the wavelength of radiation, which effectively minimized the wave impedance.

The system equipment for the SCR-1 stellarator was managed by the National Instruments for data acquisition and management. It employs a PXIe module that includes thirty-two digital input and output channels in addition to an external power supply and industrial-grade logic level. Two core instrumentation LabVIEW modules are essential within the system: one for maintaining optimal vacuum conditions through pressure regulation during the initial operational phase and the other for initiating SCR-1 systems for plasma discharge\cite{Asenjo2018}.

\begin{figure}[h!]
    \centering
    \includegraphics[width=0.9\linewidth]{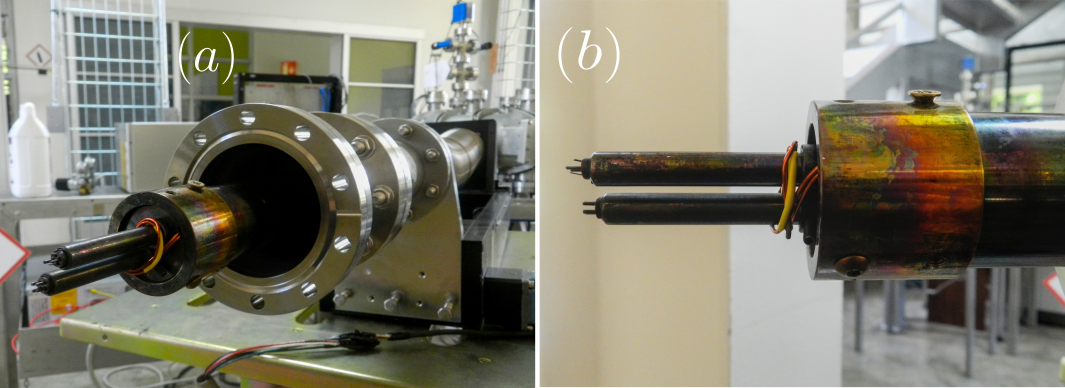}
    \caption{(a) Positioning system and (b) head of Langmuir probe designed for the SCR-1 stellarator }
    \label{fig:sonda}
\end{figure}

The plasma diagnostic method implemented in SCR-1 was the Langmuir probe. It is composed of three primary component systems: the probe head, positioning system, and data acquisition system, two of which are shown in Figure \ref{fig:sonda}. The probe head was constructed using eight cylindrical tungsten electrodes: six had a diameter of \SI{1.00}{\mm}, and two had a diameter of \SI{0.050}{\mm}. All of these electrodes feature an exposed tip length of \SI{3.5}{\mm}. The positioning system is equipped with a gate valve to isolate the vacuum vessel from the probe head. This system incorporates a linear movement mechanism that enables the displacement of a specialized steel tube designed for high-vacuum pressures. The probe tip connector was linked to PXIe, with a resistor inserted between them to protect the equipment from voltage spikes. Additionally, it utilizes the SMU source for dual functions: providing the floating potential necessary for electrode power, and collecting both ionic and electronic currents from the plasma.

\subsection{SCR-1 plasma discharge process}\label{subsubsec:download_process}
%Methodology
The plasma discharge in SCR-1 has three stages \cite{Asenjo2018}:
\begin{enumerate}
    \item Start-up: the pressure module is activated to regulate the working-base pressure.
    \item Shoot sequence: plasma discharge occurs. The following systems were activated in that order: gas injection, magnetic confinement, microwave launch, and plasma diagnostic measurements.
    \item Shutdown: the systems referenced in the preceding stage are deactivated, leaving only the vacuum chamber operational in preparation for stage (ii). To finish the experimental phase, a sequential shutdown procedure for the pumps was started, restoring the base pressure within the SCR-1 vacuum vessel to atmospheric pressure.
\end{enumerate}

%A series of plasma discharge processes were carried out according to the operational protocol for the SCR-1 stellarator in conjunction with a single Langmuir probe operation.
To elucidate the procedure, the initial parameter set for plasma discharge number 947 is summarized in Table \ref{table:par_descarga_scr1}.
\begin{table}[!ht]
\caption{\label{table:par_descarga_scr1}Initial parameters of the plasma discharge 947 for the SCR-1 stellarator.}
\begin{indented}
%\centering
\item[] \begin{tabular}{@{}cc}
\br    
Parameter & Magnitude \\
\mr
Power input $\left( \si{\kilo\watt} \right)$ & \num{3.000} \\
Gas flow $\left( \si{SCCM} \right)$ & \num{20} \\
Bank battery voltage  $\left( \si{\volt} \right)$ & \num{127.8} \\
Base presurre  $\left( \si{\torr} \right)$ & \num{4e-5} \\
%Radial probe tip position $\left( \si{\metre} \right)$ & \num{0.270} \\
\br
\end{tabular}
\end{indented}
\end{table}

\begin{figure}[!ht]
    \centering
    \includegraphics[width=1.0\linewidth]{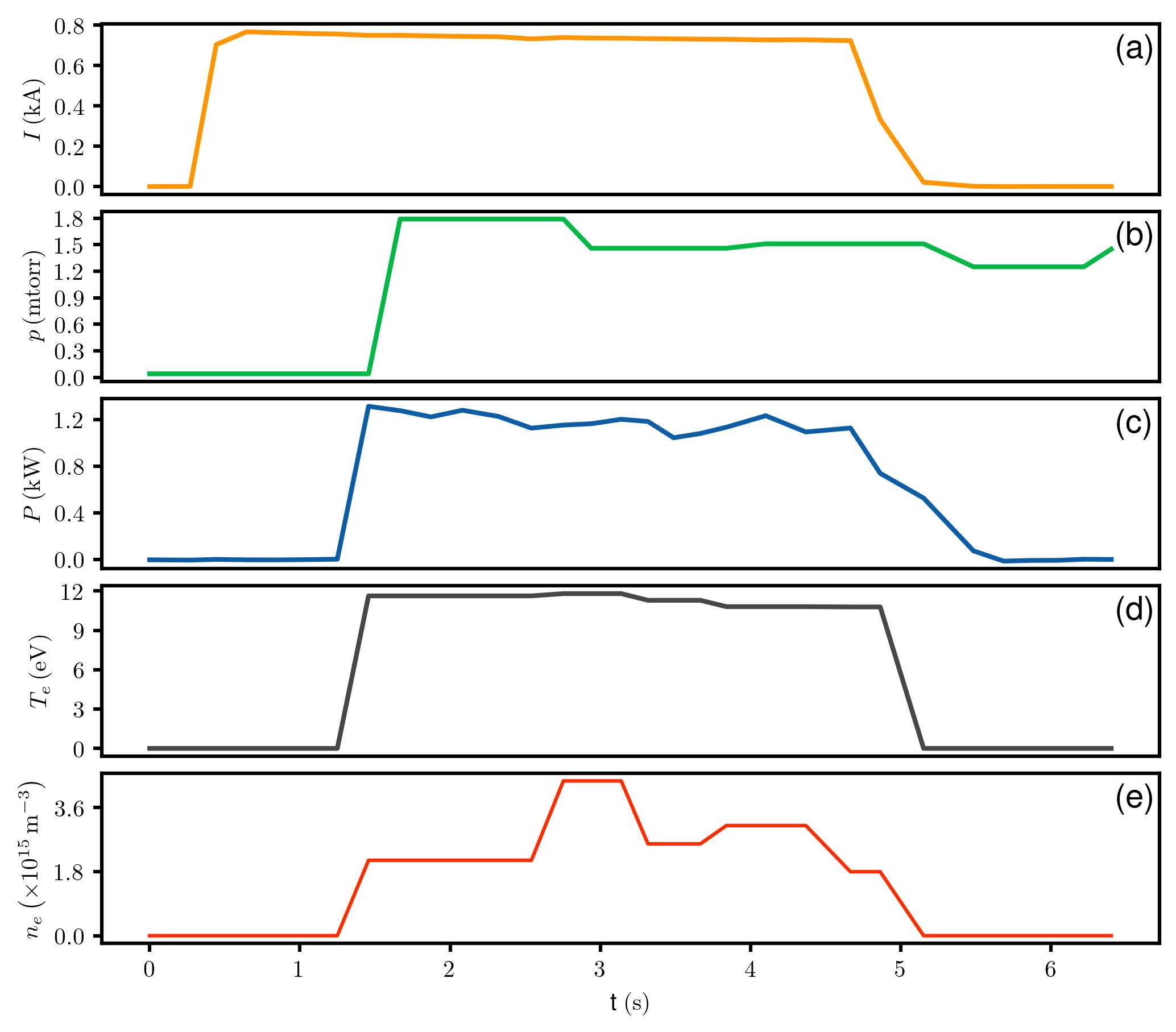}
    \caption{Evolution of key plasma discharge parameters in the SCR-1 stellarator: (a) electrical current in the coils, (b) pressure inside the vacuum vessel, (c) absorbed plasma power, (d) electron temperature, and (e) electron density.}
    \label{fig:descarga_scr1}
\end{figure}

The temporal evolution of the control system parameters for 947 plasma discharge is illustrated in Figure \ref{fig:descarga_scr1}. The electrical current in the coils is averaged at \SI{4265.6}{\ampere} per coil to create the vacuum magnetic field. The rise occurred simultaneously with gas injection, which increased the vacuum vessel pressure. Simultaneously, the magnetron was turned on to achieve the magnetic field strength required for confinement and the ECRH system was activated. During an effective discharge period of approximately \SI{4.1}{\second}, the ECRH system operated with an average absorbed power of \SI{1.149}{\kilo\watt}. The power level remained relatively steady and then decreased gradually. This observation was confirmed by the electron temperature measurements, which also remained relatively stable. The line electron density exhibited a notable peak followed by a significant decrease, possibly linked to electron-ion recombination or escape from magnetic confinement.

\section{Single Langmuir probe measurements of electron density and electron temperature radial profiles} \label{sec:SLP}

\subsection{Limit electron density estimation using power balance} \label{subsec:ne_limite}
%Metodology
The power required to maintain a low-temperature plasma discharge can be estimated by considering the overall power and the particle balance of the discharge. According to SCR-1 plasma parameters in \cite{Coto2020}, the plasma is low-temperature hydrogen plasma. To estimate the maximum electron density reached given a fixed value of electron temperature, the methodology proposed in \cite{Lechte2002} was adapted to consider the expected electron temperature range in the SCR-1. 

The power and particle balance scheme accounted for all relevant sources and losses, including absorbed power, ionization, excitation, electron-ion and electron-neutral collisions, and recombination. This estimation involves calculating the parameters specified in Table \ref{table:parameters_ne} as follows: the mean power absorbed by the plasma denotes the operational power used in the SCR-1 discharge. The plasma volume was calculated based on the MHD equilibrium results in Section \ref{sec:mhd}. The neutral density was estimated from the mass flow of the injected hydrogen. The ratio of the plasma edge electron temperature to the plasma core electron temperature was estimated from the average of this parameter using a single Langmuir probe at different discharges. It was assumed that particle diffusivity was significantly greater than thermal diffusivity. Finally, the ionization and electron-neutral impact rates for hydrogen over an electron temperature range of \SI{0.0}{\electronvolt} to \SI{20.0}{\electronvolt} were calculated.
\begin{table}[ht!]
\caption{\label{table:parameters_ne}Parameters of the confinement device and plasma of the SCR-1 for the calculation of the electronic limiting density.}
\begin{indented}
%\centering
\item[] \begin{tabular}{@{}cc}
\br    
Parameter & Magnitude \\
\mr
Absorbed power (\si{\kilo\watt}) & \num{1.149} \\
Plasma volume (\si{\meter^3}) &  \num{0.007812} \\
Neutral density ( \si{\meter^{-3}}) & \num{6.38E19} \\
\br
\end{tabular}
\end{indented}
\end{table}

\begin{figure}[h!]
     \centering
     \includegraphics[width=0.65\textwidth]{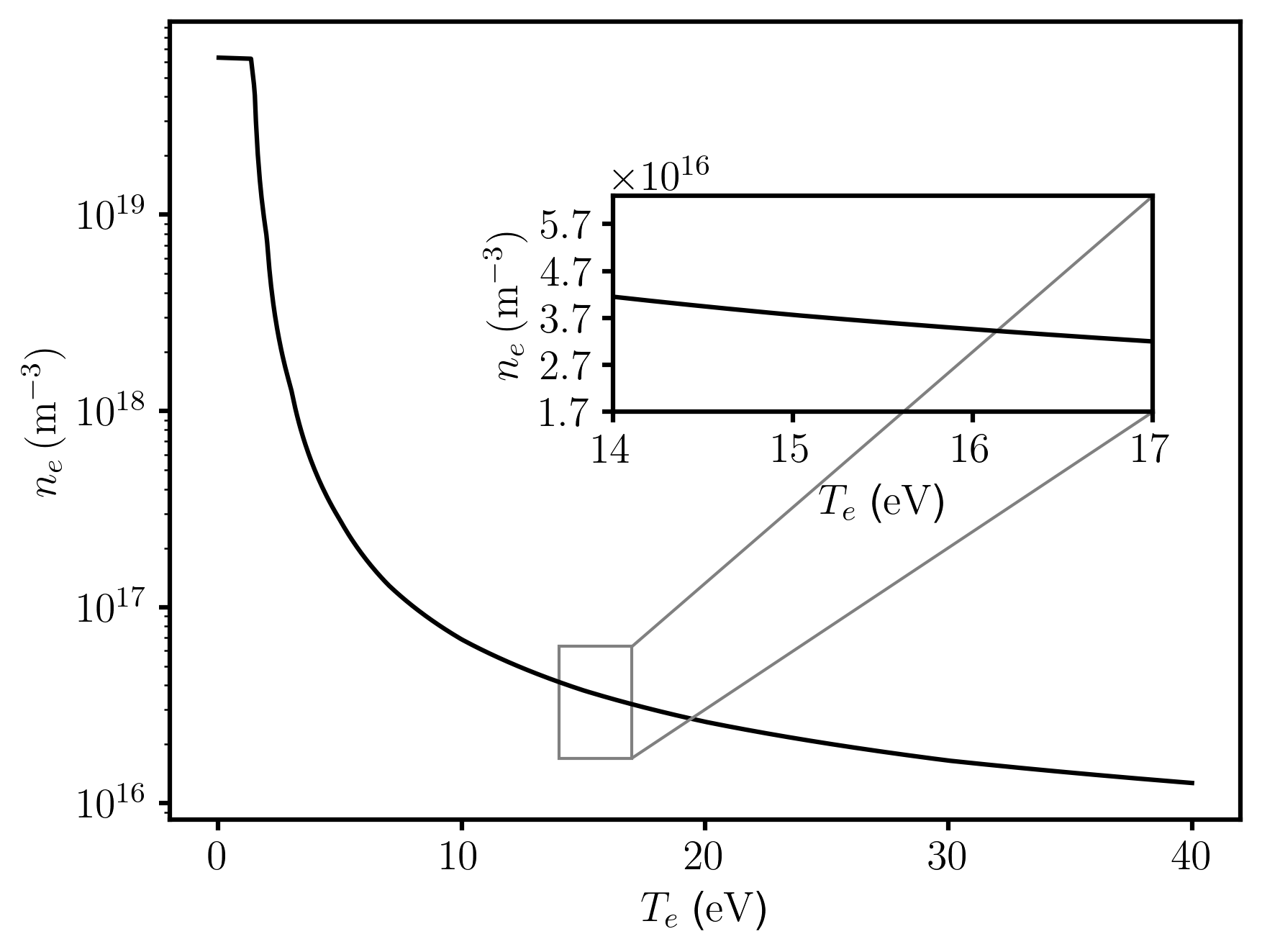}
     \caption{Maximum electron density as a function of electron temperature for the SCR-1 plasma.}
     \label{fig:ne0_vs_te}
\end{figure}

Figure \ref{fig:ne0_vs_te} shows the correlation between the highest electron density and electron temperature. The electron density decreased with increasing electron temperature. As outlined in \cite{buttenschon2018, krause2002}, this outcome was due to the high density of neutral atoms in the SCR-1 plasma, leading to significant heat losses during the ionization process, as electron-neutral collisions result in the release of energy from electromagnetic waves.

\subsection{Langmuir probe measurement of electron temperature and electron density}
\begin{figure}
    \centering
    \includegraphics[width=0.7\textwidth]{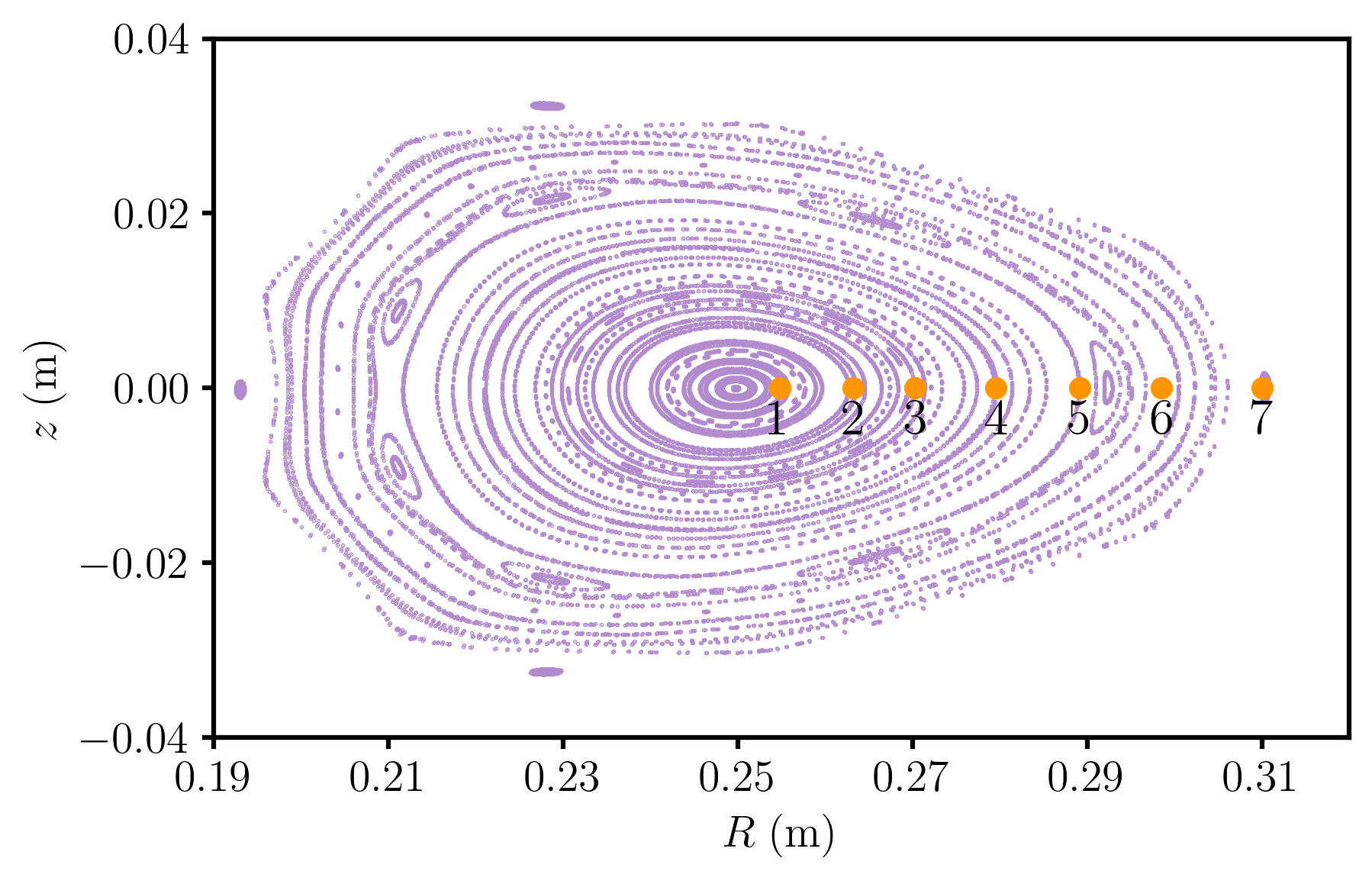}
    \caption{Positions \SI{0.254}{\metre}, \SI{0.263}{\metre}, \SI{0.270}{\metre}, \SI{0.279}{\metre}, \SI{0.289}{\metre}, \SI{0.298}{\metre}, and \SI{0.310}{\metre}, labels from 1 to 7, in the radial direction for current and voltage collection with the single Langmuir probe in the SCR-1 plasma. The vertical position of the probe was always maintained at $z = \SI{0.0}{\metre}$.}
    \label{fig:SF_medicion_langmuir}
\end{figure}

%Metodology
Electron density and electron temperature measurements for the SCR-1 plasma were conducted using a specifically designed single Langmuir probe, as described in Section \ref{sec:SCR1}. The procedure involves positioning the probe within the port corresponding to the magnetic flux surface at zero toroidal degrees and aligning it with the SCR-1 port. Subsequently, seven distinct radial positions were selected and measured relative to the major radius, as shown in Figure \ref{fig:SF_medicion_langmuir}. 

%A ruler attached to the Langmuir probe's positioning system was used to precisely determine the location of the probe within the vacuum vessel during the experimental measurement process. This determination relies on a conversion factor that relates the vacuum position to the markings on the ruler.

Once the probe was placed in the required position, the LabView-designed software was configured to collect the current and voltage values for the current-voltage characteristic curve. The software is configured using the initial parameters listed in Table \ref{table:parametros_sonda}.
%\begin{comment}
%\end{comment}
\begin{table}[ht!]
\caption{\label{table:parametros_sonda}Parameters for the initial configuration of the single Langmuir probe in the SCR-1 plasma discharge process.}
\begin{indented}
%\centering
\item[] \begin{tabular}{@{}cc}
\br    
Specification & Magnitude \\
\mr
Floating Voltage (\si{\volt}) & \num{-40} to \num{70} \\
Data Acquisition Rate (\si{samples\per\second}) & \num{40} \\
Floating Voltage Frequency (\si{\hertz}) & \num{2} \\
\br
\end{tabular}
\end{indented}
\end{table}

As pointed out in \cite{conde2011}, the ratio of the Langmuir probe tip length to the Debye length plays a role in shaping the ``knee'' and subsequently influences the plasma parameter magnitudes.  According to \cite{Popov2012}, under these conditions, it is necessary for the electron mean free path to be significantly longer than the probe tip size. Assuming a Maxwellian distribution and following the methodology presented in \cite{Livadiotis2019}, the electron mean free path for the SCR-1 plasma was estimated to be approximately \SI{e3}{\metre}. This value satisfies the condition when compared to the dimensions specified in section \ref{sec:SCR1}. The Larmor radius ($r_L$) ranged from \SI{0.128}{\mm} to \SI{0.372}{\mm}, and this range was estimated based on the electron temperature data at various radial positions and the magnitude of the corresponding magnetic field. These values closely align with the radius of the Langmuir probe tip ($r_p$), necessitating the interpretation of the current-voltage characteristic curve within the framework of an electron diffusion model in a highly magnetized plasma. This implies that electrons in the SCR-1 plasma have limited mobility in the direction perpendicular to the magnetic field, leading to lower electron and ion saturation currents, as estimated by the first-derivative method \cite{stanojevic1994}. 

Consequently, adjustments were made to the model and the corrected plasma potential was computed following the methodology detailed in \cite{bhuva2019}. The electron saturation current was obtained using Equation \ref{ec:Ise}, which combines the measured electron current ($I_{se}^{*}$) calculated by interpolating the floating potential value in the the current-voltage characteristic curve and multiplying this current by a scaling factor.
\begin{eqnarray}\label{ec:Ise}
    I_{se} = I_{se}^{*}\left( 1 + \frac{\pi}{8} \frac{r_{p}}{r_{L}} \right)
\end{eqnarray}
This correction allowed the electron saturation current for the SCR-1 plasma to increase by an order of magnitude at positions 1, 2, 3, and 4 in Figure \ref{fig:SF_medicion_langmuir}, where the largest Larmor radii occurred.

Finally, according to \cite{Usoltceva2018}, under the aforementioned conditions, measurements must be supplemented by a method to calculate the effective collection area used to estimate the electron density. This is because, when the probe tip is not perpendicular to the magnetic field, the particle flux is reduced, defining an effective collection area that is smaller than the total area. Points 1, 2, 3, and 4 in Figure \ref{fig:SF_medicion_langmuir} exhibit the most significant deviations from the perpendicularity between the magnetic field vectors and the probe tips. A reduction in \SI{7}{\percent}-\SI{8}{\percent} in the effective collection area of the Langmuir probe was observed towards the center of the plasma relative to the total probe area. Consequently, the electron density slightly increased by more than an order of magnitude.

To measure the electron density in plasma of these plasma lengths, it is imperative to assess whether the probe tips are influenced while consistently immersed in the plasma. The Debye length varied from \SI{0.71}{\mm} to \SI{0.26}{\mm} at the radial positions where measurements were conducted. Consequently, the screening effect was extended to \SI{3.6}{\mm} (equivalent to five Debye lengths) using the Langmuir probe tip. This result indicates that the plasma sheath formed in a region with a radius twenty-one times smaller than the minimum plasma radius \cite{marks2011}. Therefore, given the significant discrepancy in dimensions, the Langmuir probe tips did not exert a substantial impact on the SCR-1 plasma.

\begin{figure}[h!]
    \centering
    \includegraphics[width=0.65\textwidth]{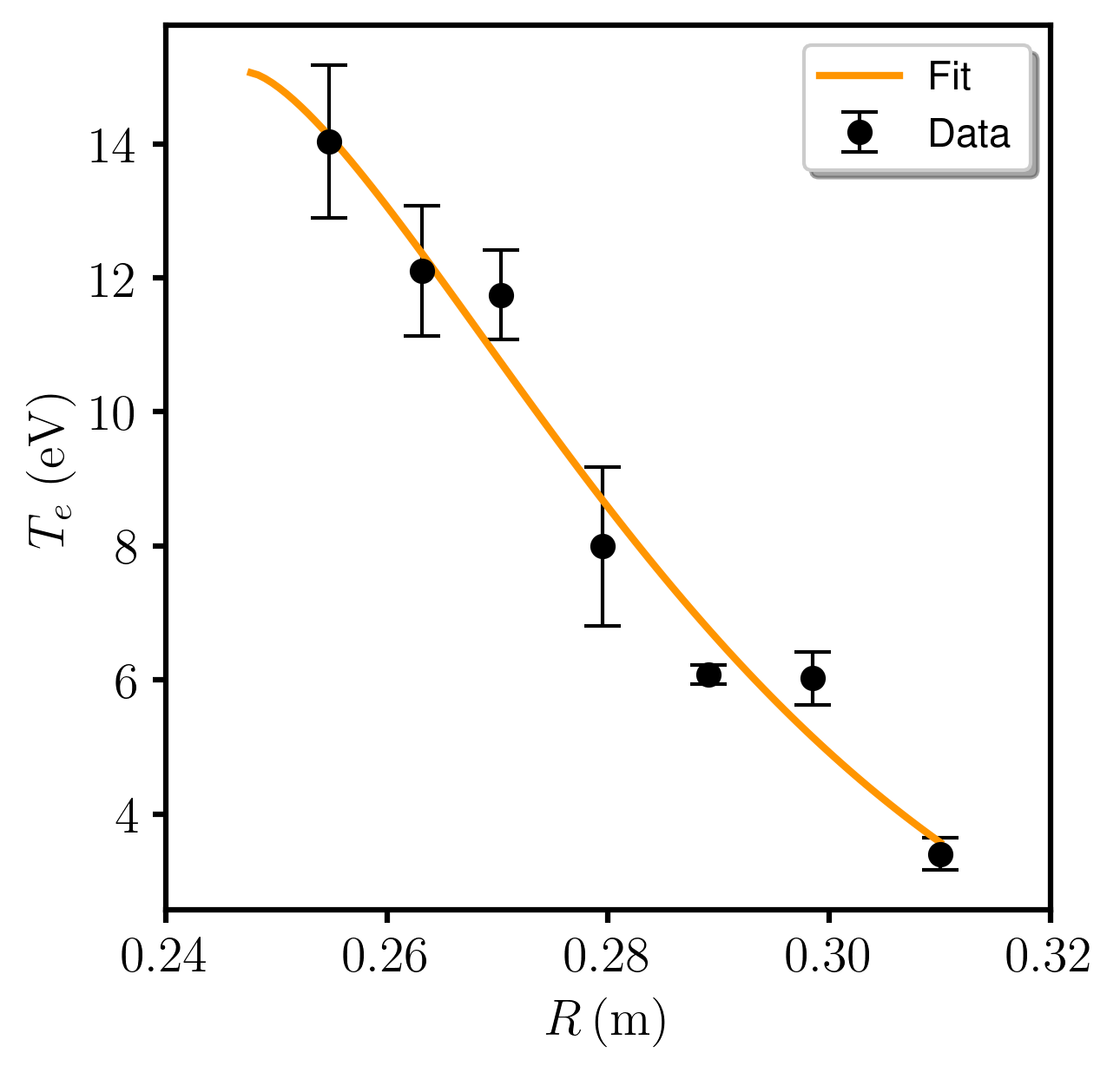}
    \caption{Measured radial profile of electron temperature}
    \label{fig:perfil_Te}
\end{figure}

\begin{figure}[h!]
    \centering
    \includegraphics[width=0.65\textwidth]{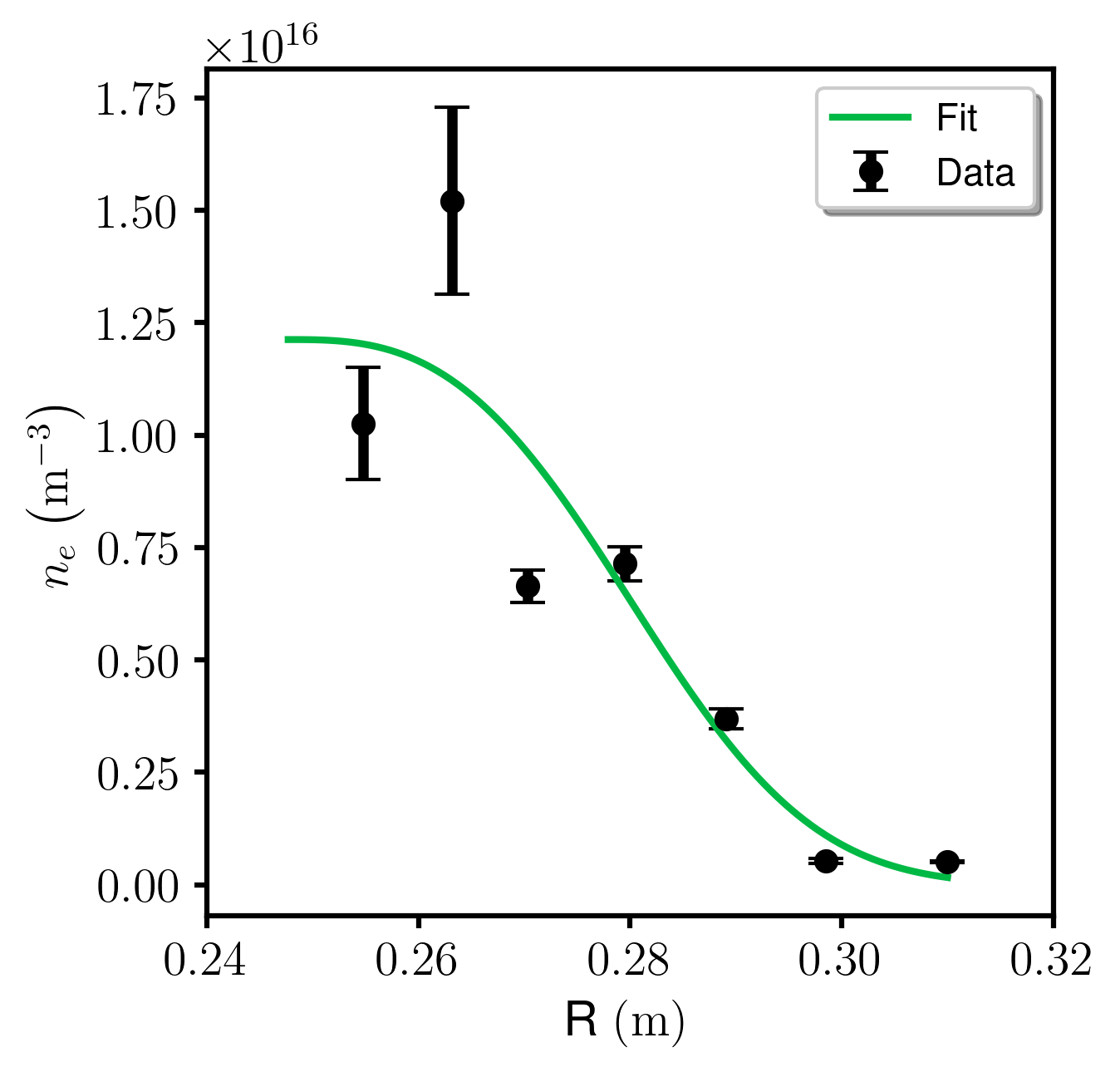}
    \caption{Measured radial profile of electron density}
    \label{fig:perfil_ne}
\end{figure}

After calculating the electron temperature and density values for each measurement point, they were fitted to a Gaussian distribution using the approach proposed in \cite{Kohn2011}. These radial profiles are presented in Figures \ref{fig:perfil_Te} and \ref{fig:perfil_ne}, and their algebraic expressions, in terms of major radius, are given by
\begin{eqnarray}
    T_{\text{e}} \left( R \right) = \SI{15.1}{\electronvolt} \exp \left( \frac{R - \SI{0.2477}{\metre}}{\SI{0.048}{\metre}} \right)^{1.4} \\
    n_{\text{e}} \left( R \right) = \SI{1.21e16}{\metre^{-3}} \exp \left( \frac{R - \SI{0.2477}{\metre}}{\SI{2.88}{\metre}} \right)^{1.57} 
\end{eqnarray}

%\vspace{-2cm}

The radial profile of the electron temperature exhibited the expected behavior for Electron Cyclotron Resonance (ECR) heating at the plasma's central axis. An increase in electron temperature was noted at radial positions closer to the magnetic axis ($R_0=\SI{0.247}{\metre}$) due to power absorption by electrons in the region where the second harmonic of the electron cyclotron rotation frequency occurs ($R=\SI{0.246}{\metre}$) \cite{prater2011}. The electron density profile exhibited the characteristic ECR profile observed in low-aspect-ratio stellarators \cite{okamura1999} and was categorized as subdense plasma \cite{stott1994, bilato2009}.

Comparing the central value of the electron density radial profile with the results in Figure \ref{fig:ne0_vs_te}, the electron density limit was not exceeded. However, the value was lower, and the standard deviations for both the electron density and electron temperature at positions near the core were smaller. A possible cause is a ``shadow'' effect in the regions near the Langmuir probe insulating supports, which could have prevented charge collection and further reduced the effective area of the probe tip. Overestimation of the electron density has been reported to be lower than or equal to \SI{10}{\percent} of their expected value \cite{Li_2023}, although high electron densities ($> \SI{e15}{\metre^{-3}}$) reduce this effect. Contamination owing to layer deposition on the probe tip is another possible cause. As reported in \cite{Deolivera2019}, variable resistance is generated, which affects the current-voltage curve for tips that are eroded or deposited with other materials.

\section{Magnetohydrodynamic Equilibrium Calculations for SCR-1}\label{sec:mhd}
\subsection{Input Parameters for VMEC}
\begin{figure}
    \centering
    \includegraphics[width=0.6\textwidth]{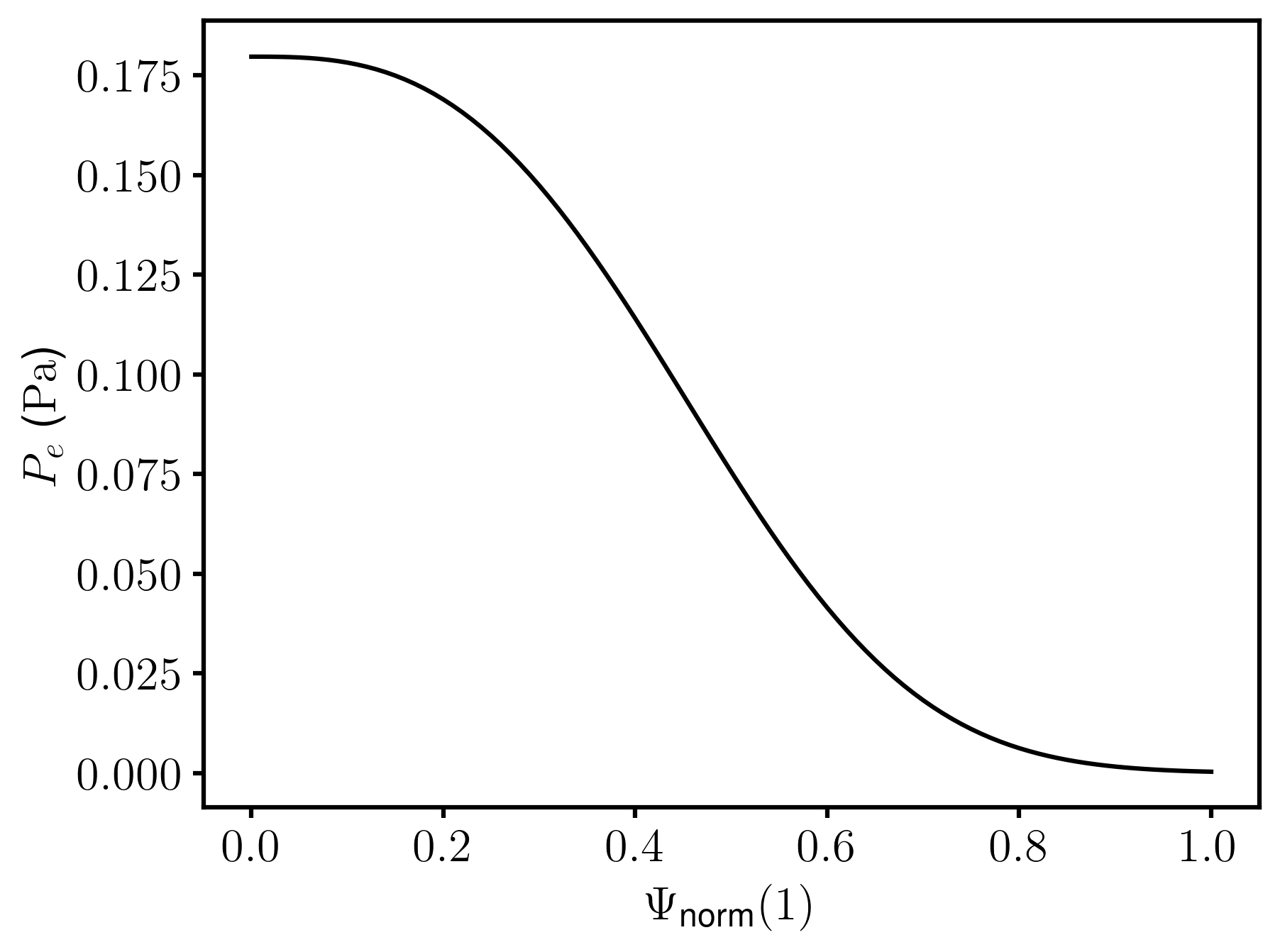}
    \caption{Radial profile of plasma-pressure}
    \label{fig:presion}
\end{figure}

%Initially, the MAKEGRID code calculates the magnitude of the magnetic field produced by the SCR-1 coils. The input parameters include the Cartesian coordinates of the \num{12} modular coils and the total current flowing through a single filament.
The magnetohydrodynamic equilibrium of the SCR-1 plasma was computed using VMEC in free-boundary mode, following a methodology similar to that in \cite{Coto2020}. In the VMEC input file available in \cite{PlasmaTEC-ITCR2023}, a new variable was introduced: the incorporation of an experimental radial electron pressure profile, defined by the following expression
\begin{equation}
    P_e = n_e T_e
\end{equation}
where $n_e$ is electron density profile shown in Figure \ref{fig:perfil_ne} and $T_e$ is electron temperature profile shown in Figure \ref{fig:perfil_Te} and shown in Figure \ref{fig:presion}.

\begin{comment}
    This parameter was incorporated through Fourier coefficients and denoted as $a_m(n)$. This adjustment was performed by fitting
\begin{equation}
    P_e = \sum_{n=0}^{10} a_m(n) \cdot \Psi^n_{norm}
\end{equation}
where $\Psi_{norm}$ is normalized magnetic flux. 
\end{comment}

\subsection{Magnetohydrodynamic Equilibrium Parameters} 
VMEC outcomes were examined using interactive notebooks obtained from PlasmaTEC-ITCR \cite{PlasmaTEC-ITCR2023} and STELLOPT \cite{Stellopt} repositories. The MHD parameters listed in Table \ref{tabla:parametros_VMEC} exhibit slight disparities when compared to the data presented in \cite{Coto2020}, which initially assumed a linear plasma-pressure profile. Geometric metrics, including major and minor radial dependent aspect ratios, plasma volume, and beta parameters, were computed. The magnetic field's pressure, which has a value significantly below one and is particularly strong in the toroidal direction, contains the plasma. This enhances plasma stability but reduces plasma confinement by amplifying derivative effects as well as restricting the attainable electron temperature and density. The dissimilarity in the beta values implies that the magnitude of the toroidal magnetic field significantly surpasses that of the poloidal magnetic field. 
\begin{table}[ht!]
\caption{\label{tabla:parametros_VMEC}Geometric and magnetic quantities from MHD calculation equilibrium}
\begin{indented}
%\centering
\item[] \begin{tabular}{@{}cc}
\br    
Parameter & Magnitude \\
\mr
Aspect ratio & \num{6.2017} \\
Plasma volume (\si{\metre^3}) & \num{0.007812} \\
Major radius (\si{\metre}) & \num{0.2478} \\
Minor radius (\si{\metre}) & \num{0.0396} \\
Total beta (\si{\percent}) & \num{0.0205} \\
Toroidal beta (\si{\percent}) & \num{1.57} \\
Poloidal beta (\si{\percent}) & \num{0.0429} \\
\br
\end{tabular}
\end{indented}
\end{table}

\begin{figure}
    \centering
    \includegraphics[width=0.6\textwidth]{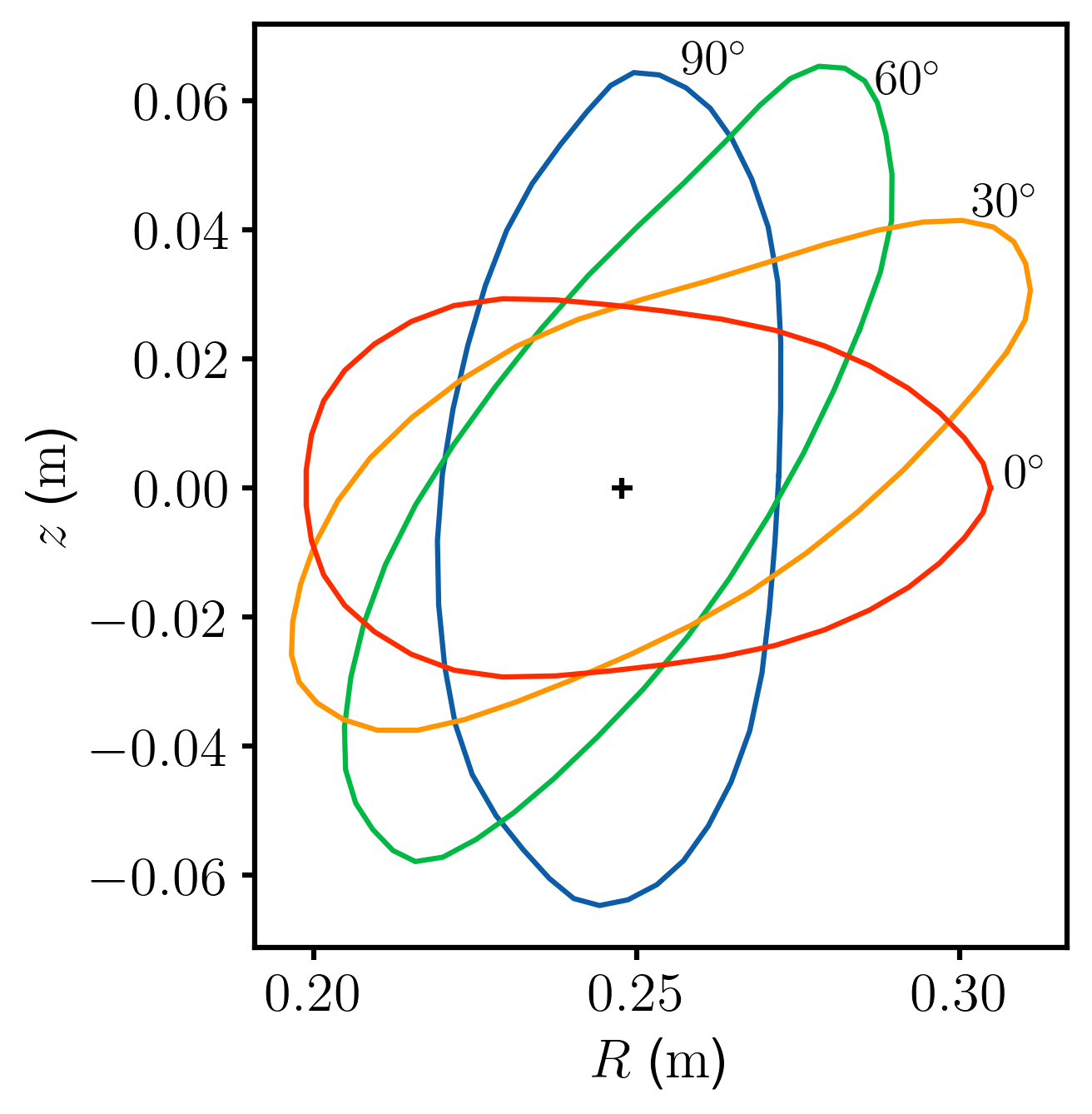}
    \caption{Last close magnetic flux surfaces for various toroidal positions in the plasma of the SCR1 stellarator.}
    \label{fig:LCMS}
\end{figure}

\begin{figure}[!h]
    \centering
    \includegraphics[width=1.0\textwidth]{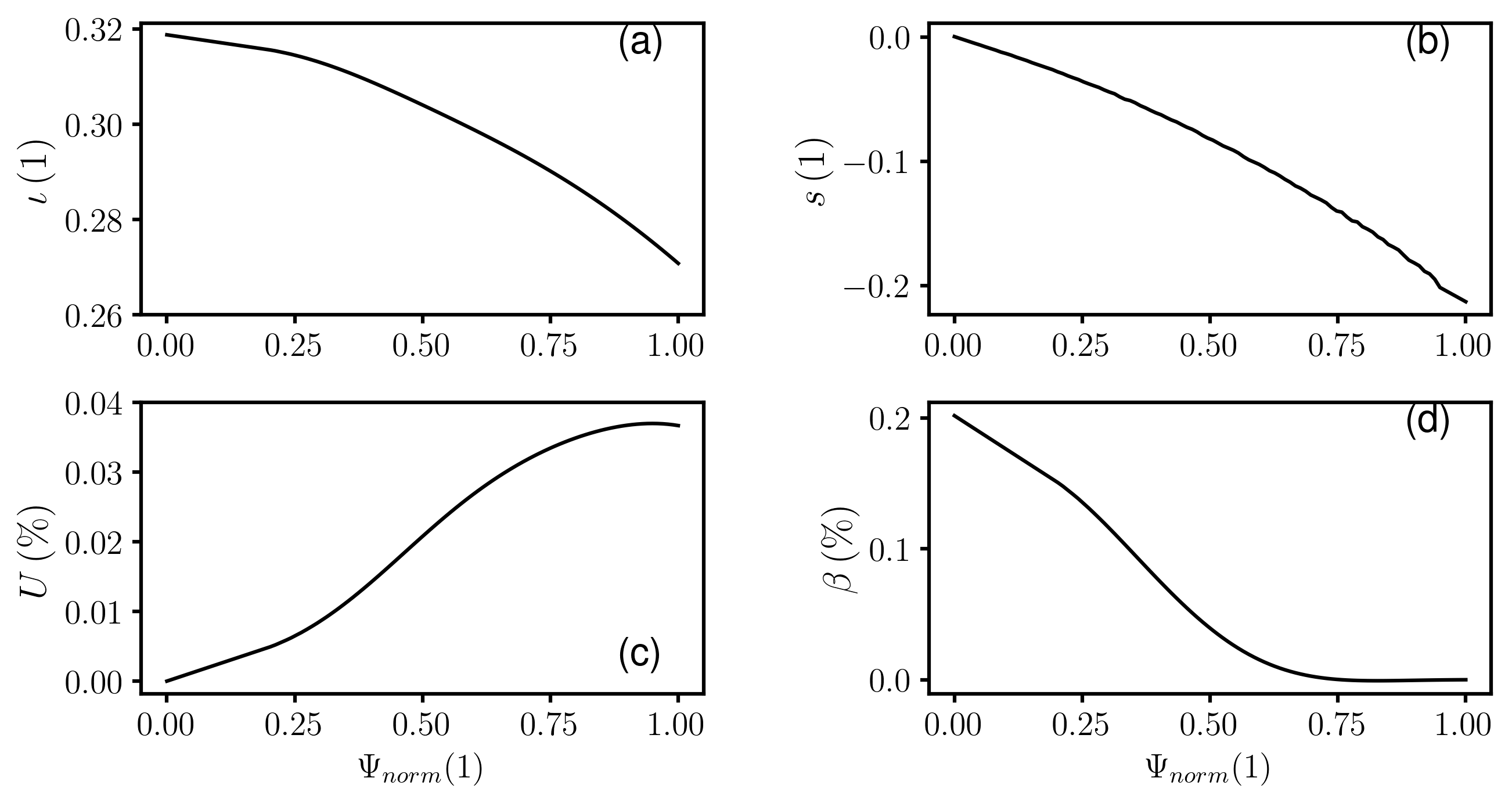}
    \caption{Radial variation of (a) iota, (b) magnetic shear, (c) magnetic well depth and (d) beta parameter from VMEC calculations at the toroidal position of \ang{0}.}
    \label{fig:variables_equilibrio}
\end{figure}

Each last closed magnetic surface  calculated at various toroidal locations, presented in Figure \ref{fig:LCMS}, preserves its elliptical shape with a length of approximately \SI{0.12}{\cm}. Figure \ref{fig:variables_equilibrio} depicts the radial profiles of iota, magnetic shear, magnetic well depth, and total beta, which are the four parameters characterizing the MHD equilibrium of the SCR-1 plasma. The iota variable decreased in the radial direction with a percentage difference of approximately \SI{18}{\percent} compared with the plasma core. This decrease reveals the presence of three rational iota values: 3/10, 2/7, and 3/11, respectively. Notably, the value of 2/7, which is situated at the position $\Psi_{norm} = \num{0.76}$ equivalent to $R = \SI{0.2911}{\metre}$, was accompanied by a significant magnetic island aligned with $z=0$ shown in Figure \ref{fig:SF_medicion_langmuir}. Magnetic shear demonstrated low and negative behavior. Consequently, charged particles subjected to magnetic curvature drift had an inversion of their radial to vertical direction, thereby remaining confined within the medium. This phenomenon contributes to plasma stability \cite{Antonsen1996}. The magnetic well depth exhibited a negative value until $R = \SI{0.298}{\metre}$, signifying the presence of a magnetic well \cite{deAguilera2015, nuhrenberg2016}. Finally, the beta parameter displayed a low average value that was approximately one order of magnitude lower than the plasma core value.

\subsection{Mercier Criterion} 
\begin{figure}[!h]
    \centering
    \includegraphics[width=0.8\textwidth]{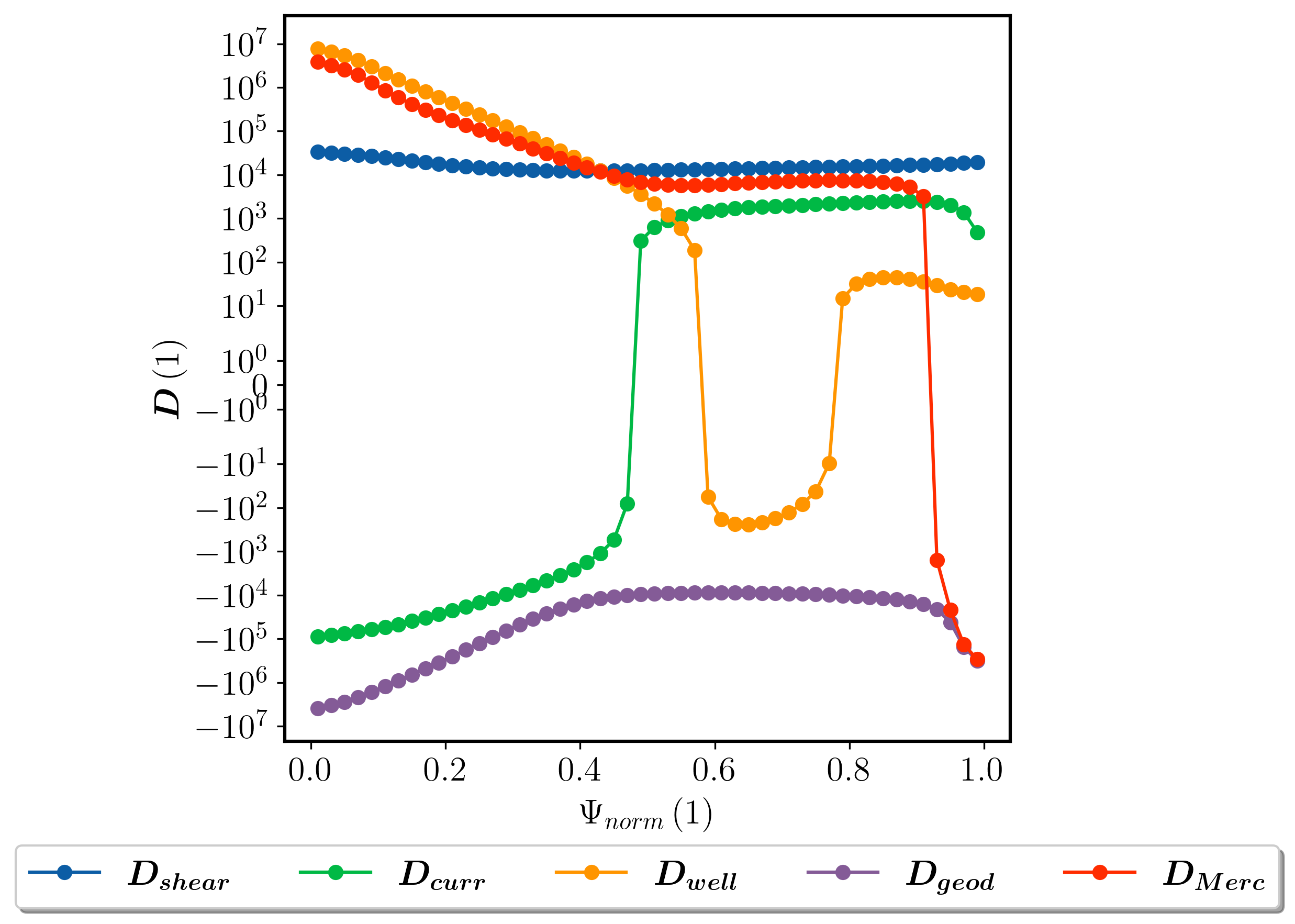}
    \caption{Variation of the magnetic shear, toroidal current, magnetic well, geodesic, and total terms for the Mercier criterion.}
    \label{fig:mercier}
\end{figure}

The SCR-1 exhibited a stable MHD equilibrium. The radial distribution of the linear stability components of the SCR-1 plasma is shown in Figure \ref{fig:mercier}. The magnetic shear component was consistently positive; however, its magnitude was significantly smaller than that of the Mercier component. The toroidal-current component undergoes a change in sign from negative to positive, which is attributed to the diamagnetic properties of the plasma. This result implies that poloidal currents generate a magnetic field that displaces the magnetic flux surfaces outward, as documented in \cite{kovrizhnykh1983, ozeki1998}. The magnetic well component is positive in regions  $\num{0.0} \leq \Psi_{norm} \leq \num{0.6}$ and $\num{0.8} \leq \Psi_{norm} \leq \num{1.0}$. The positive curvature of the magnetic flux surfaces is advantageous for their stability, as highlighted in \cite{Landreman2020}. Among the various components, the magnetic well term contributes the most to the total Mercier term. However, the geodesic term consistently exhibits a negative value owing to the presence of the Pfirsch-Schlüter current. This destabilizing behavior was previously reported by \cite{harada2003}. The total Mercier term becomes negative at a radial position of \SI{0.297}{\metre} near a magnetic island located at \SI{0.292}{\metre}. Similar results of this behavior near the plasma edge were reported for the TJ-II stellarator \cite{deAguilera2015}.

\section{Microwave heating scenarios} \label{sec:IPF-FMDC} 

\subsection{Full wave code IPF-FMDC}
%Introducción al código
The selection of the IPF-FMDC 3D full wave code for the simulation of various heating scenarios within the SCR-1 is based on its success in multiple magnetic confinement devices such as TJ-K, TJ-II, RF-X, Pegasus Toroidal Experiment, CNT and MAST \cite{kohn2008, bilato2009, Kohn2011, Hammond2018, kohn-seeman2023}. The code aims to visually represent the transmission of electromagnetic waves through the plasma in order to identify the phases of the O-X-B and X-B conversion processes.

%Parámetros de entrada
The input files for the IPF-FDMC code consist of a two-dimensional matrix based on $R$ and $z$ coordinates, formatted according to the HDF5 standards. These files contain normalized values of the electron density relative to the cutoff density of the ordinary mode and the magnitude of the three Cartesian components of the confining magnetic field relative to the resonant magnetic field for ECR heating.

%Radiación
The simulation of radiation in the IPF-FDMC code utilizes a ray with a K Gaussian spectrum \cite{kohn2008}. This method entails the superposition of a single component of the electric field oriented along a single grid line with the amplitude of the electric field varying uniformly perpendicular to it. The incident electric field is given by 
\begin{eqnarray}
    E_z \left( x \right) = A \left( t \right) E_0 \exp \left[  -\frac{\left(x-x_0 \right)^2}{w_0^2} \right]
\end{eqnarray}
where $x$ is the position of the beam, $x_0$ is the center of the ray, $w_0$ is the beam waist, $E_0$ is the amplitude of the electric field in a vacuum, and $A(t)$ is a function that takes values between 0 and 1 for different oscillation periods.

\subsection{Selected parameters for plasma heating scenarios} \label{subsec:sphs_scr1}
Microwave heating scenarios within the SCR-1 were exclusively simulated through the O-X-B mechanism conversion process. This was owing to the curvature of the conversion zones surrounding the plasma core, which failed to align with the highest magnetic field strength. Consequently, the launch of radiation from high-field zones is impossible for a possible X-B mechanism.

The parameters for the heating scenarios in the SCR-1 were chosen after analyzing the dependencies of the O-X conversion percentage, as reported in \cite{kohn2008,guo2017}. These parameters include the toroidal and poloidal angles, beam width, electron density, and magnetic field modulus. The angles in the toroidal and poloidal directions were selected based on the geometry of the inner port of the SCR-1 vacuum vessel. All scenarios were simulated using an incident radiation frequency of \SI{2.45}{\giga\hertz} with polarization in the ordinary mode.

%The maximum angular aperture measured from the center of the plasma was approximately \ang{26}, and was a circular aperture. The range of $\left[ \ang{-26}, \ang{26} \right]$ was defined poloidally, and $\left[ \ang{-25}, \ang{25} \right]$. 
An electron density input file is formalized using the radial electron density profile specified in Figure \ref{fig:perfil_ne}. The power and particle balancing discussed in Section \ref{sec:SCR1} were employed to create an overdense plasma by adjusting the absorbed power while maintaining a constant electron density core temperature, similar to \cite{buttenschon2018}. Three electron density ratios with respect to the cut-off electron density from the ordinary mode were selected to represent the limits of the current SCR-1 heating system. The ray-tracing code BS-Solctra was employed to generate magnetic field files \cite{Jimenez2019}. Three magnetic field ratios were selected, both above and below the value of the electronic cyclotron heating in the second harmonic.The IPF-FDMC full-wave code was executed on the CICIMA-HPC \cite{cicimahpc} cluster to ensure the efficient and timely completion of the simulation.

%All scenarios were simulated using an incident radiation frequency of \SI{2.45}{\giga\hertz} with polarization in the ordinary mode. 

\subsection{Scenario 1}
\begin{figure}[!h] %Modificar tamaño para que se vea mejor cada gráfico
    \centering
    \includegraphics[width=1.0\linewidth]{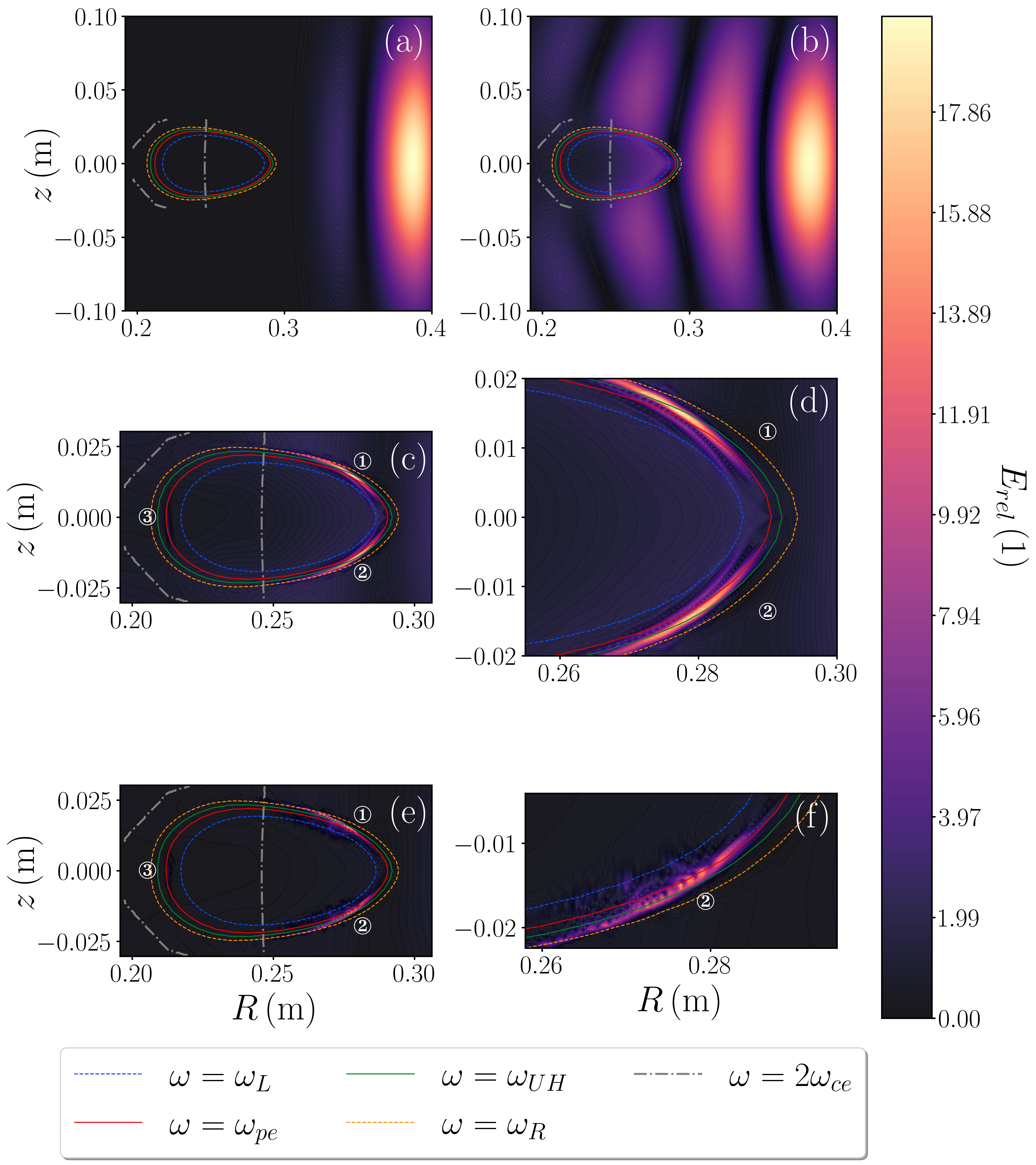} %mecanismoOXB
    \caption{The O-X-B conversion mechanism simulated with the IPF-FDMC full-wave code, visualized from variations in the electric field normalized to its maximum value.}
    \label{fig:mecanismoOXB}
\end{figure}

The present scenario aims to examine the increase in electron density within the plasma while maintaining the magnetic field configurations employed in the SCR-1 experimental procedure.  Figure \ref{fig:mecanismoOXB} illustrates the O-X-B conversion scheme in the single-pass mode, based on the absolute value of the electric field magnitude of the radiation and the conversion regions for the SCR-1 plasma. This scheme was applied to the specific cases of $n_e/n_{e\,cut}=\num{2.14}$ and $B/B_{ce}=\num{1.00}$.

%Figure \ref{fig:mecanismoOXB} presents the O-X-B conversion process in a single-pass mode, utilizing the absolute value of the electric field magnitude of the radiation, normalized to its maximum value, and the conversion regions specific to the SCR-1 plasma. This scheme was applied to the specific cases of $n_e/n_{e\,cut}=\num{2.14}$ and $B/B_{ce}=\num{1.00}$. 

In Figure \ref{fig:mecanismoOXB}.(a), electromagnetic waves with a planar wavefront propagate towards the ordinary mode cut-off region in the plasma. Figure \ref{fig:mecanismoOXB}.(b) depicts the O-X (ordinary to extraordinary slow) conversion process: the waves in the ordinary mode encounter the ordinary mode cutoff region, resulting in light diffraction around this region and the emergence of interference patterns. After the \num{12} oscillation periods, three distinct spatial regions were identified, as shown in Figure \ref{fig:mecanismoOXB}.(c) and labeled with numbers \ding{172}, \ding{173}, and \ding{174}, wherein the electric-field oscillations were absorbed near the zone between the ordinary mode cut-off and upper hybrid frequency region. Consequently, a fraction of the radiation is converted to the extraordinary mode, whereas the remaining fraction is directed towards the dampers at the edges. The simulation revealed a standing-wave pattern during this period. The incident waves approached the O-X conversion region, as shown in Figure \ref{fig:mecanismoOXB}.(d), they adapted to the shape of the ordinary mode cut-off for O-X conversion, with wavelengths on the order of the Larmor radius owing to a reduced group velocity \cite{urban2011}. However, some light aberrations were maintained, indicating that the light did not reach the cutoff point. After \num{25} periods, the intensity of the electric-field oscillations decreased compared with the previous stage (Figure \ref{fig:mecanismoOXB}.(e)), and propagated perpendicular to the magnetic field beyond the conversion zone, as shown in Figure \ref{fig:mecanismoOXB}.(f). The latter description is associated with the propagation of electronic Bernstein waves towards the center of the plasma, in agreement with \cite{kohn2011_1, arefiev2016, bilato2009}.

\begin{figure}[!h]
    \centering
    \includegraphics[width=0.8\linewidth]{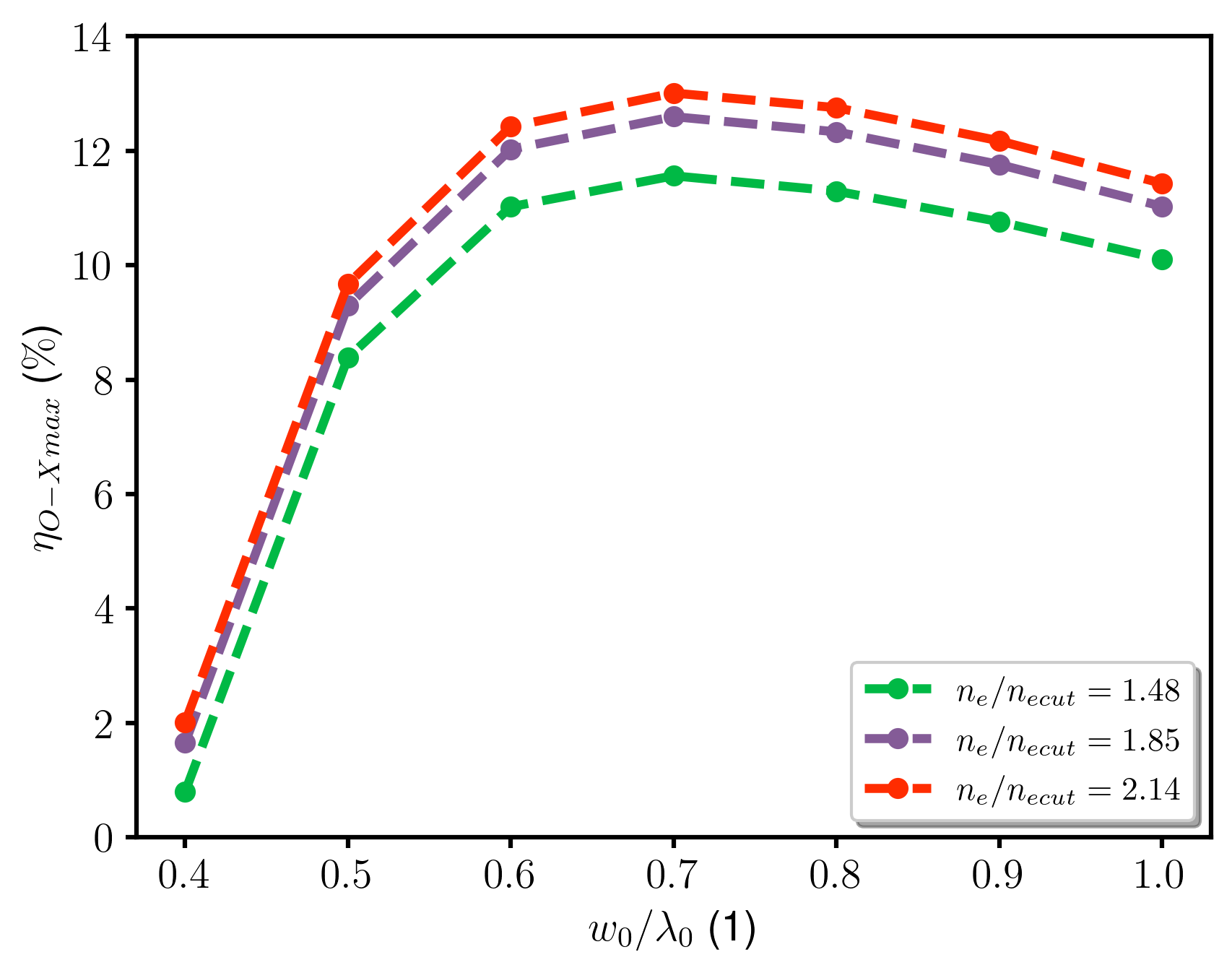}
    \caption{Maximum O-X conversion percentage as a function of the incident beam waist ratio with the incident radiation wavelength ($\lambda_0$).}
    \label{fig:etavsancho_experimental}
\end{figure}

Figure \ref{fig:etavsancho_experimental} demonstrates the correlation between the maximum O-X conversion percentage and the ratio of the incident beam width to its wavelength for three different electron density ratios. An electron density ratio of \num{2.14} yielded a maximum O-X conversion percentage of \SI{13.0}{\percent}. The optimal incident beam width that achieved a balance between the divergence of the light beam and the poloidal curvature was determined to be $\num{0.7}\lambda_0$.  

To support the increase in the O-X conversion percentage with a higher electron density ratio, the trend of each parameter in Table \ref{tabla:k0lnyRcurvox} for region \ding{173} was presented. These parameters exhibit similar characteristics in regions \ding{172} and \ding{174}. 

The decrease in the electron density resulted in the displacement of the ordinary mode cutoff region, which was always accompanied by a significantly pronounced electron density gradient. This gradient causes the normalized electron density scale length ($k_0L_n$) to remain below one, which is detrimental to the O-X percentage in the SCR-1 plasma, as reported by \cite{guo2017}. The triangularity of this zone was \num{0.4}, leading to greater diffraction of electromagnetic waves when interacting with the plasma, owing to a drastic change in the refractive index over a short distance compared to the dimensions of the plasma \cite{bilato2009}. However, an increase in the O-X conversion percentage was observed when a lower value of $k_0L_n$ was used, as demonstrated in \cite{pochelon2007,kohn-seeman2023}. There was an increase in the O-X conversion rate when a lower value of this normalized scale length was obtained for certain ranges of this variable, as reported in \cite{pochelon2007,kohn-seeman2023}. These studies found that an increase in $k_0L_n$ led to a greater dependence on launch angles. As a result, due to not matching the optimal angle for the SCR-1, the O-X conversion percentage decreased. A reduced $L_n$ does not necessarily result in an improvement in the O-X conversion. This occurs because if the electronic density gradient changes slowly in the region near the cut-off of the ordinary mode, the last peak of the electric field of the incident radiation significantly increases in amplitude, reaching a higher value before decreasing to match the solution of the slow extraordinary mode. This occurs when the optimal launch angle is achieved \cite{kohn2008}. A larger ordinary mode cutoff curvature radius indicates a better match to the radiation wavefront \cite{kohn2008, bilato2009}. This aspect plays a crucial role in the difference between the aforementioned optimal beam widths because a pronounced curvature introduces an additional phase variation that widens the spectrum of the Gaussian beam and results in a decrease in the O-X conversion percentage \cite{shalashov2014}. 
The O-X conversion percentage decreased as the $k_0L_B$ values decreased for the SCR-1 plasma location. The findings presented in the citation by \cite{guo2017} altered the trend of the magnetic field scale length. It might possibly be connected to a magnetic beach, which occurs when the electromagnetic waves of the plasma are in an area where the magnetic field undergoes substantial variations. The fluctuation in the magnetic field can impact electromagnetic waves. Finally, the characteristic coupling scale length decreases with the electron density ratio, as in \cite{Bin2013}.
\begin{table}[ht!]
\caption{\label{tabla:k0lnyRcurvox}Relevant parameters for the analysis of the OX conversion mechanism for the SCR-1 plasma in region \ding{173} of scenario 1.}
\begin{indented}
%\centering
\item[] \begin{tabular}{@{}cccccc}
\br    
$n_e/n_{ecut}$ & $\eta_{O-Xmax} (\si{\percent})$  & $k_0L_{n}(1)$  & $k_0L_{B} (1)$ & $R_{O} $ (\si{\cm}) & $L_{\nabla}$ (\si{\cm}) \\ 
\mr
\num{1.48} & \num{11.6} &  \num{0.528} & \num{8.76} &  \num{3.16} &  \num{1.00} \\
\num{1.85} & \num{12.6} & \num{0.399} & \num{8.79} &  \num{3.41}  &  \num{0.84} \\
\num{2.14} & \num{13.0} & \num{0.287} & \num{9.21} &  \num{3.47}  &   \num{0.75} \\
\br
\end{tabular}
\end{indented}
\end{table}

\begin{figure}[!h]
    \centering
    \includegraphics[width=0.8\linewidth]{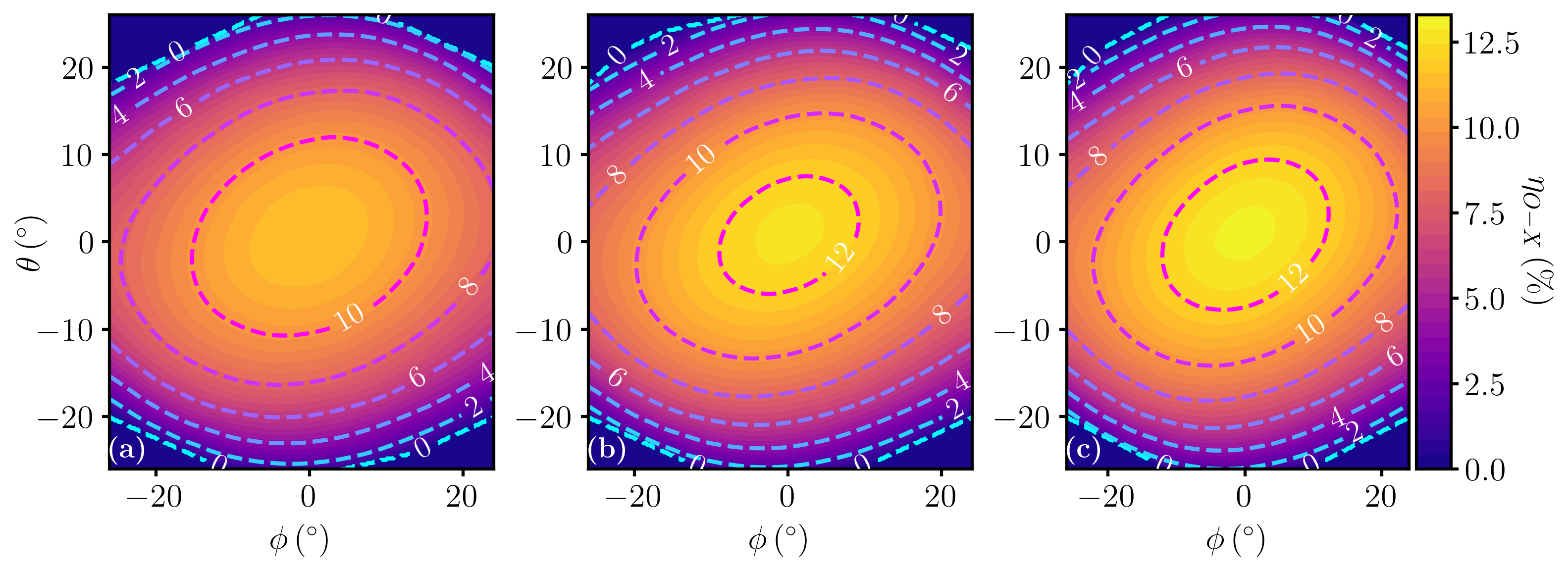}
    \caption{Angular conversion window of the O-X conversion percentage for the electron density ratios: (a) $\num{1.48}n_{e \, cut}$,  (b) $\num{1.85}n_{e\, cut}$; and (c) $\num{2.14}n_{e\, cut}$.}
    \label{fig:ventana_angular}
\end{figure}

Figure \ref{fig:ventana_angular} illustrates the angular windows for the maximum O-X conversion percentage for the three selected electron-density ratios as a function of the poloidal and toroidal angles. As expected, the widest angular window is shown in Figure \ref{fig:ventana_angular}.(c), which corresponds to the lowest value of $k_0L_n$ \cite{kohn2011_1}. In this scenario, the poloidal and toroidal angles range from \ang{0} to \ang{1} without reaching the \SI{100}{\percent} O-X conversion percentage. Similar results were observed in \cite{ikeda2008, guo2017}, indicating that there is no fixed optimal polarization or direction for radiation when the plasma length scale is comparable to the wavelength of the electromagnetic waves (i.e., when $k_0L_n < 1$). The plasma of the stellarator SCR-1 exhibited a negligible dependence on the angle, resulting in the highest conversion percentage when the radiation traveled perpendicular to the magnetic field.

\subsection{Scenario 2}
In this scenario, the electron density ratio of \num{2.14} was employed. A variation of \SI{5.0}{\percent} in the magnitude of the magnetic field was implemented to adjust the ratio between the magnetic field magnitude and the second-harmonic resonant magnetic field. A change in the magnitude of the confining magnetic field has a significant impact on the radial profile of the plasma electron density \cite{Dinklage2005}. However, Scenario 2 focused solely on the effect of displacing the resonant electron cyclotron heating region in relation to the plasma core on the O-X conversion percentage.

\begin{figure}[!h]
    \centering
    \includegraphics[width=0.8\linewidth]{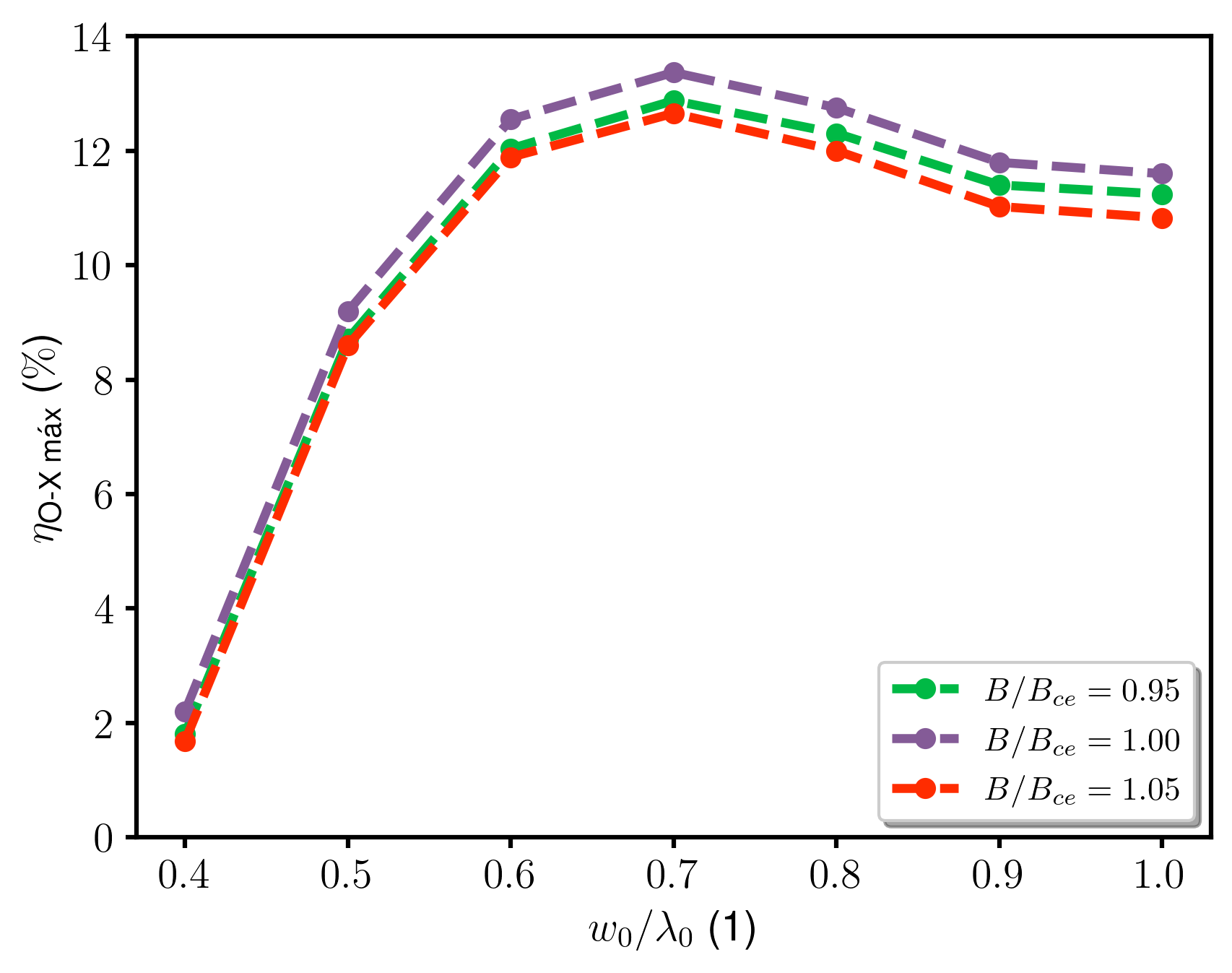}
    \caption{Maximum O-X conversion percentage as a function of the incident beam width (beam waist) obtained using the IPF-FDMC full-wave code for three ratios of magnetic field and resonant magnetic field magnitudes.} 
    \label{fig:etavsancho_B}
\end{figure}

The maximum O-X conversion percentage was achieved when the magnetic field was set to $\num{1.00}B_{ce}$, as illustrated in Figure \ref{fig:etavsancho_B}. The relevant parameters explaining the variation in the O-X conversion percentage are listed in Table \ref{tabla:k0lnlbr_B}. These parameters exhibited behavior consistent with that observed in Scenario 1, where the characteristic scale length of the electron density decreased. Consequently, no significant changes in the O-X conversion percentage were observed for the SCR-1 plasma when the resonance position of the ECR heating was varied.
\begin{table}[ht!]
\caption{\label{tabla:k0lnlbr_B}Relevant parameters for the analysis of the OX conversion mechanism for the SCR-1 plasma in region \ding{173} of scenario 2.}
\begin{indented}
%\centering
\item[] \begin{tabular}{@{}cccccc}
\br    
$B/B_{ce}$  & $\eta_{\text{O-X max}} (\si{\percent})$  & $k_0L_{n}(1)$  & $k_0L_{B} (1)$ & $R_{O} $ (\si{\cm}) & $L_{\nabla}$ (\si{\cm}) \\ 
\mr
\num{0.95} & \num{12.5}  & \num{0.263} & \num{8.60} & \num{3.47} &  \num{0.74}  \\
\num{1.00} & \num{13.0}  & \num{0.244} & \num{9.30} & \num{3.47} &  \num{0.75}  \\
\num{1.05} & \num{12.3}  & \num{0.260} & \num{8.41} & \num{3.47} &  \num{0.76}  \\
\br
\end{tabular}
\end{indented}
\end{table}

\begin{figure}[!h]
    \centering
    \includegraphics[width=1\linewidth]{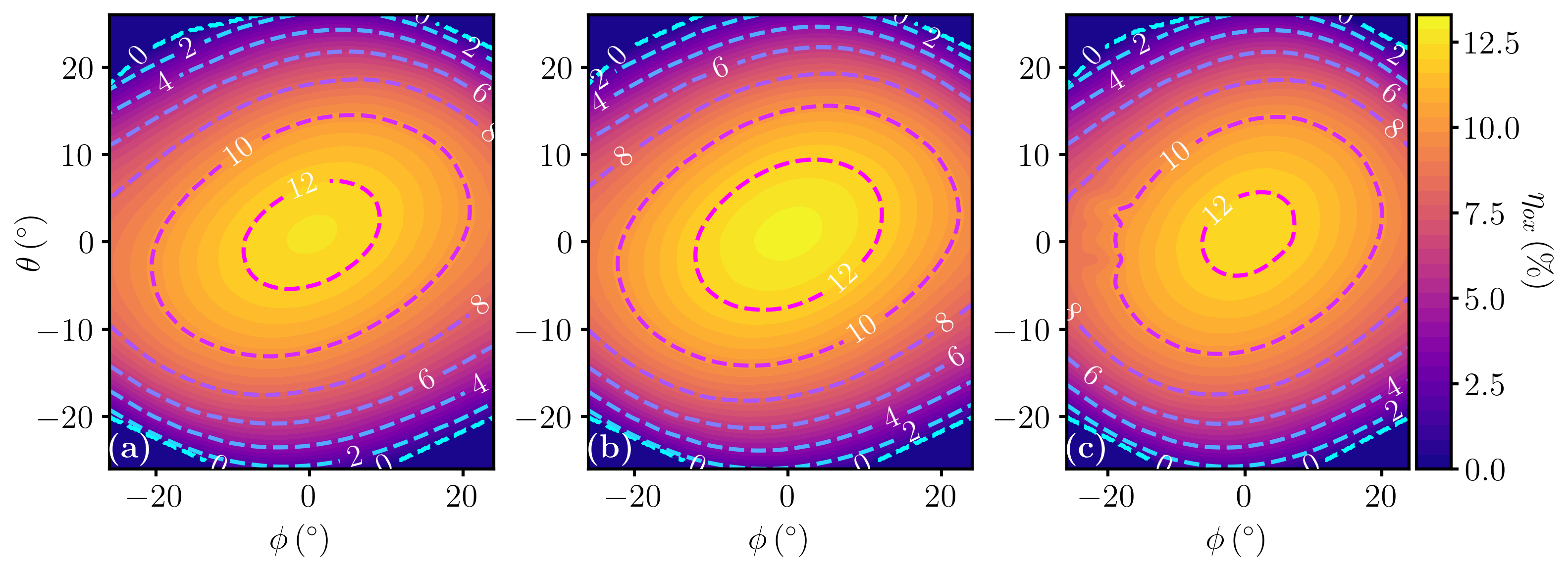}
    \caption{Angular conversion window of the O-X conversion percentage for the magnetic field magnitude ratios: (a) $\num{0.95}B_{ce}$; (b) $\num{1.00}B_{ce}$; and (c) $\num{1.05}B_{ce}$.}
    \label{fig:ventana_angular_B}
\end{figure}

The angular windows, as shown in Figure \ref{fig:ventana_angular_B}, exhibited a similar trend to that observed in Scenario 1, with no clear dependence on the angles in the toroidal and poloidal directions.

\subsection{Scenario 3}
Scenario 3 encompasses full-wave simulations with multiple reflections in 2D geometry. The geometry of the vacuum vessel was adjusted according to the characteristics presented in Table \ref{table:characteristics_SCR1} with an open waveguide coupled to the vacuum vessel, which had a diameter of \SI{7.29}{\cm}. Scenario 3 incorporates the three electron density ratios from Scenario 1 and a magnetic field magnitude of $\num{1.0}B_{ce}$. The toroidal angle was varied from \ang{0} to \ang{26} in steps of \ang{1}, whereas the poloidal angle was fixed at \ang{0}. A fixed beam width of $\num{0.7}\lambda_0$ is set.

\begin{figure}[!h]
    \centering
    \includegraphics[width=1.0\linewidth]{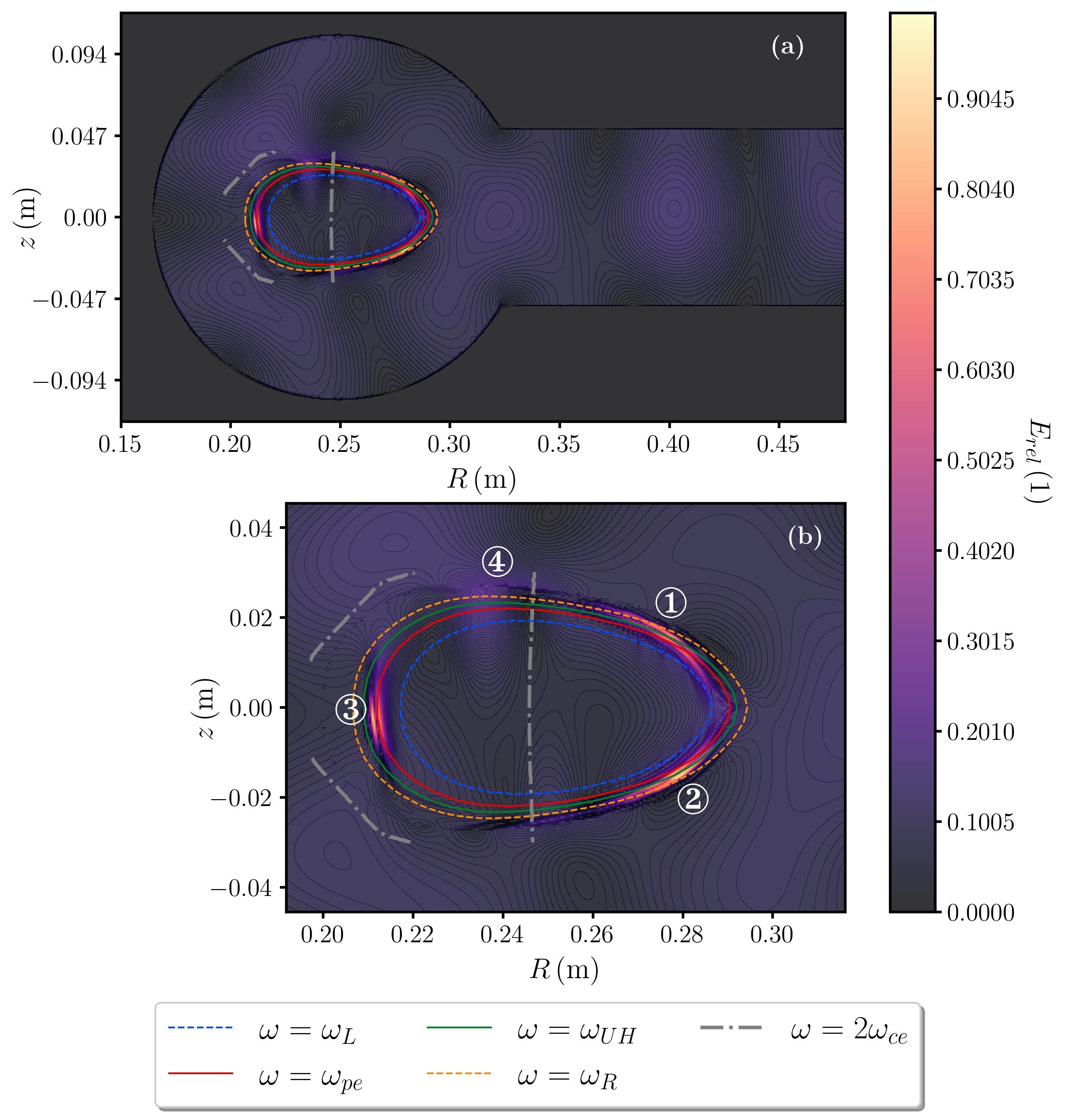}
    \caption{Variations in the electric field normalized from its maximum value of incident radiation in the $Rz$ space after completing \num{18} oscillations (steady state) for the electronic density ratio \num{2.18} in scenario 3. (b) Zooming in on the location of the conversion regions where four O-X conversion zones are labeled.}
    \label{fig:mecanismoOXB_camara}
\end{figure}

Figure \ref{fig:mecanismoOXB_camara} shows the standing waves with an absolute value of the electric field of radiation inside the SCR-1 vacuum vessel. The launch angle was \ang{15} in the toroidal direction. The three conversion zones are shown in Figure \ref{fig:mecanismoOXB}.(c), were identified. In this scenario, the aforementioned zones covered a larger conversion area, particularly in zone \ding{174}, despite having slightly higher values of $k_0L_B$, $k_0L_n$, and $L_\nabla$ than zone \ding{173}, as illustrated in Table \ref{tabla:parametros_n214_camara}. Additionally, a new conversion zone, labeled \ding{175}, was observed, characterized by a low electric field magnitude but with the advantage of being in close proximity to the electron cyclotron resonance heating region.
\begin{table}[ht!]
\caption{\label{tabla:parametros_n214_camara}Relevant parameters for the analysis of the OX conversion mechanism for the O-X-B plasma in the SCR-1 in region \ding{174} of scenario 3.}
\begin{indented}
%\centering
\item[] \begin{tabular}{@{}cc}
\br    
Parameter & Magnitude \\ 
\mr
$k_0L_{n}(1)$ & \num{0.420}\\   
$k_0L_{B} (1)$ & \num{9.84} \\
$L_\nabla (\si{\cm})$ & \num{0.94} \\
\br
\end{tabular}
\end{indented}
\end{table}

\begin{figure}[!h] %Modificar
    \centering
    \includegraphics[width=0.7\linewidth]{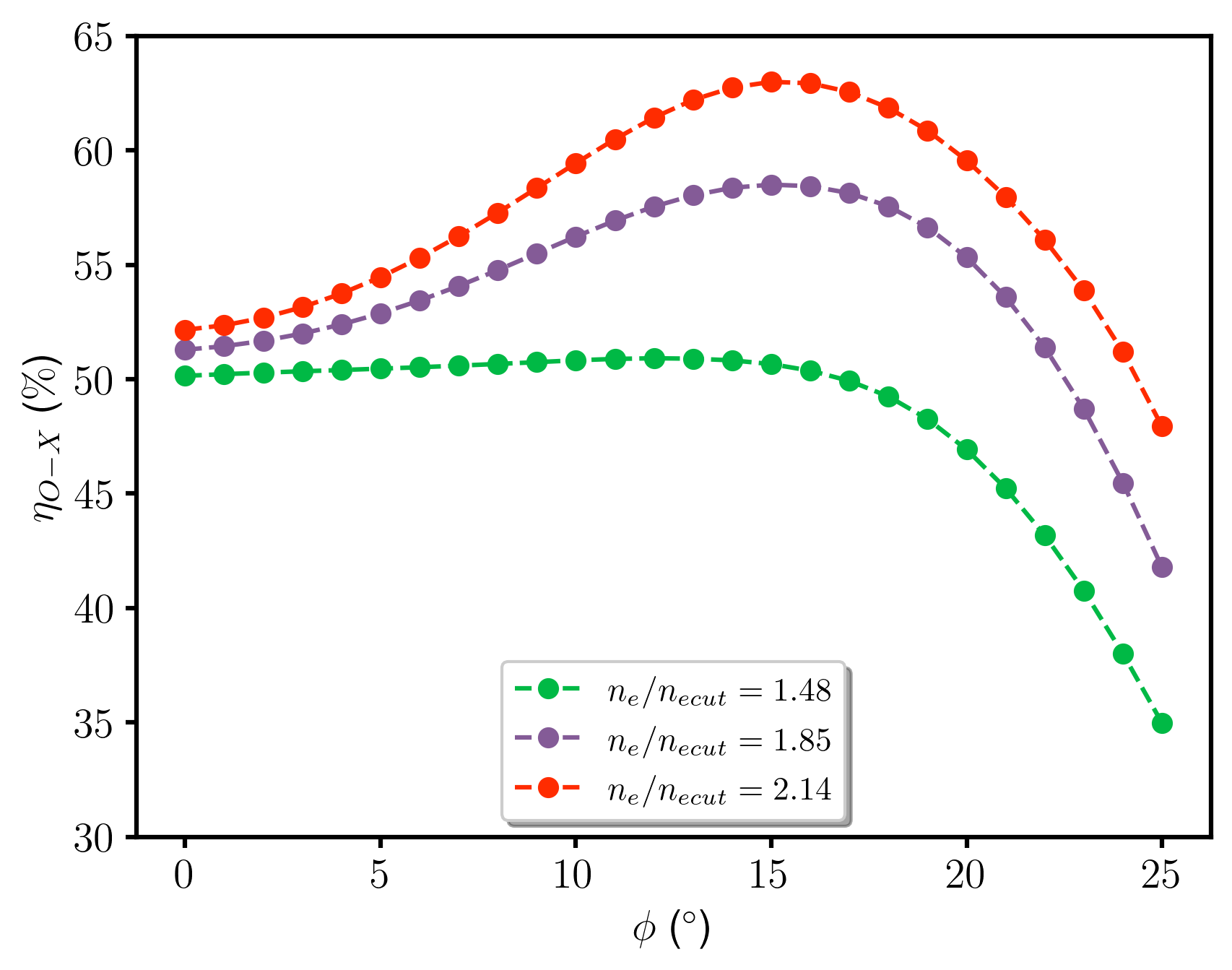}
    \caption{Maximum O-X conversion percentage as a function of the toroidal angle obtained for three electronic density and cutoff electronic density ratios with $w_0=\num{0.7}\lambda_0$, $B/B_{ce}=\num{1.0}$, and $\theta = \ang{0}$ from scenario 3. The dashed line only connects the points.}
    \label{fig:eta_vs_phi_camara}
\end{figure}

Figure \ref{fig:eta_vs_phi_camara} illustrates the variation in the O-X conversion percentage as a function of the electron-density ratio for different toroidal angles. In comparison to Scenario 1, the maximum conversion was observed at a nonzero toroidal angle, resulting in \SI{63}{\percent}. No significant improvements were observed when the simulation was performed for the different beam widths. The evolution of each electron density ratio agrees with previous reports \cite{kohn2011_1, Nagasaki2002} and Scenario 1. 

According to previous studies \cite{kohn2008, Kohn2011, Hammond2018}, the reflection of radiation inside the vessel was facilitated by the high reflectivity of aluminum (\SI{90}{\percent} - \SI{95}{\percent}) at the frequency of the incident radiation \cite{callister2020}, resulting in enhanced absorption in the regions labeled as ordinary mode cutoff. This enhanced absorption was attributed to the presence of an optimal wavefront, characterized by an optimal parallel component of the wave vector that achieved maximum O-X conversion \cite{khusainov2018}. During the interval spanning from the eighth to tenth period, a considerable portion of the unabsorbed radiation was redirected back into the waveguide, resulting in interference between the incoming and outgoing electromagnetic waves. This interference poses a threat to the experimental equipment.

\subsection{Non-resonant absorption mechanisms for Bernstein waves} \label{subsec:LimSXB}
Motivated by the findings in Scenarios 1, 2, and 3, the non-resonant absorption mechanisms of electron Bernstein waves in the SCR-1 were investigated and analyzed, particularly in the vicinity of the upper hybrid frequency. The three most significant absorption mechanisms for the SCR-1 plasma with input parameters $B/B_{ce} = \num{1.0}$ and $n_e/n_{ecut} = \num{2.14}$ are described for scenario 3.

\begin{figure}[!h] 
    \centering
    \includegraphics[width=0.7\linewidth]{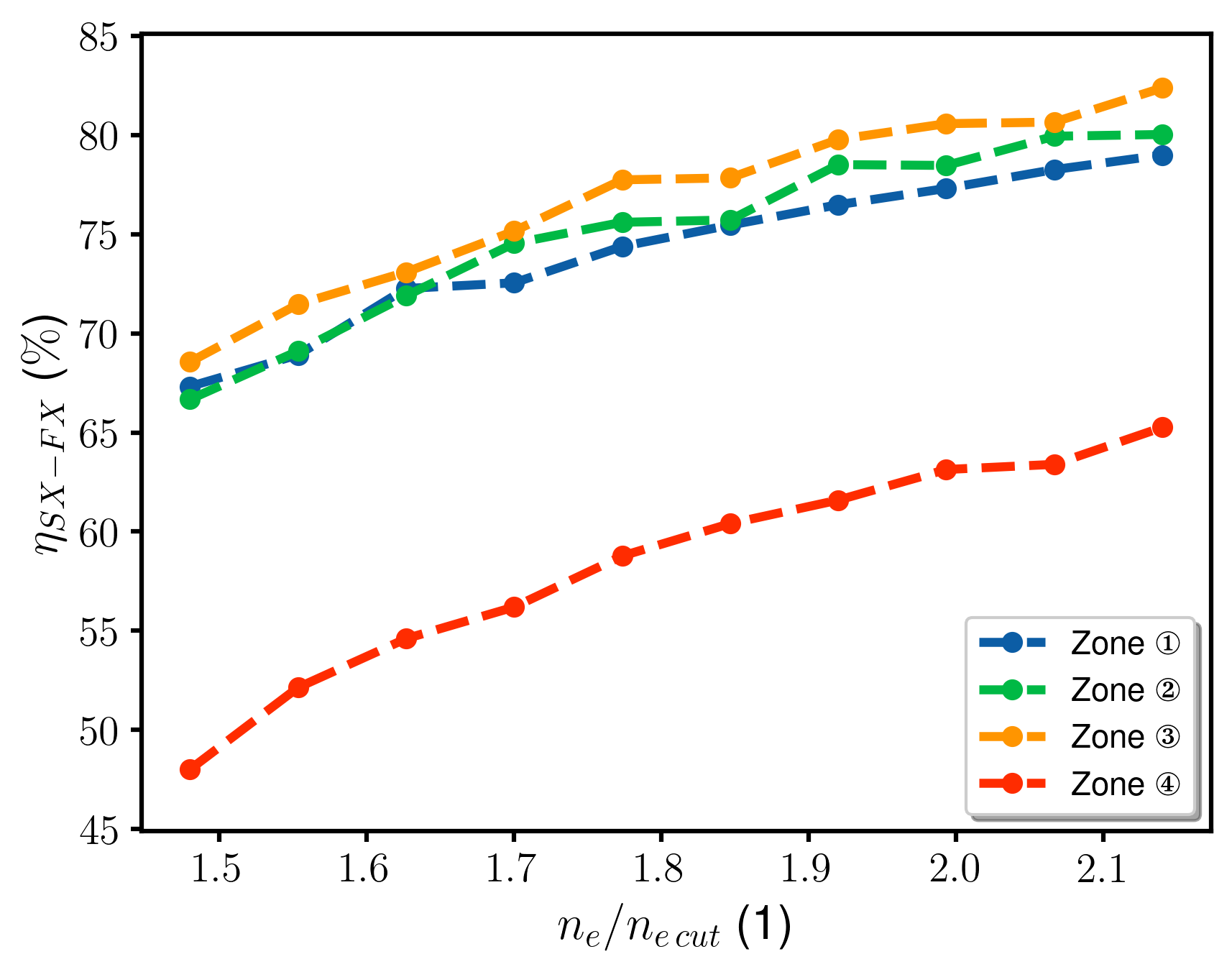}
    \caption{Parameter $\eta_{SX-FX}$ for the values of the electronic density ratio and cutoff electronic density within the defined range for the SCR-1 plasma in each conversion region identified in the heating scenarios.} 
    \label{fig:tau_bvsne}
\end{figure}

The SX-FX conversion was computed for different electron-density ratios, as detailed in \cite{ram2000}. The results are shown in Figure \ref{fig:tau_bvsne} for each of the identified O-X-B conversion regions. This observed pattern was associated with a reduction in $k_0L_n$ as the electron density increased \cite{guo2017}.

Table \ref{tabla:parametros_prop_bernstein} presents the relevant variables for analyzing the damping of electron Bernstein waves in the regions near the X-B conversion. These variables were calculated as follows:
\begin{enumerate}
    \item The wavelength of the electric field near the UHR zone ($\lambda_E$) was estimated based on the variations in the electric field obtained from simulations using the full-wave code IPF-FDMC. It was assumed that the frequency of the incident radiation remained constant during the O-X-B conversion mechanism.
    \item The ion-electron collision frequency ($\nu_{ie}$) was calculated using the method described in \cite{richardson2019}, which considers the electron temperature, electron density, and Coulomb logarithm.
    \item The electron-neutral collision frequency ($\nu_{en}$) was estimated using \cite{wang2021}, with the data reported from the cross-section in the IAEA database named ALADDIN \cite{ALADDIN}.
\end{enumerate}

The ratio of the ion-electron collision frequency to the frequency of the incident radiation was slightly greater than \num{e-5} for all three selected electron density ratios. Therefore, these collisions do not significantly impede the propagation of the electron Bernstein waves. A small value of this ratio indicates that collisions occur infrequently and have a minimal impact on wave propagation \cite{ikeda2008, Hammond2018, diem2018}.

The Bernstein mode was determined to have traveled from a radial position of \SI{0.279}{\meter} to a radial position of \SI{0.121}{\meter} in Figure \ref{fig:mecanismoOXB_camara} after undergoing \num{164} oscillations, at which point collisions between the electrons and neutral atoms would likely have occurred. This enabled synchronization between the Bernstein mode and the electrons, thus allowing the waves to continue propagating to the ECR heating zone. Therefore, the electron-neutral collision frequency is not a significant mechanism for damping the B mode.
\begin{table}[ht!]
\caption{\label{tabla:parametros_prop_bernstein}Relevant parameters in the analysis of the propagation of electronic Bernstein waves for region \ding{173}. The trend of these parameters was similar for the four absorption regions.}
\begin{indented}
%\centering
\item[] \begin{tabular}{@{}cccc}
\br    
$n_e/n_{e\,corte}$ & $\lambda_E/r_L$  & $\nu_{ei} \, (\times \SI{e4}{\hertz} )$ & $\nu_{en} \, (\times \SI{e7}{\hertz})$  \\ 
\mr
\num{1.48} & \num{2.2} &  \num{6.48} & \num{1.45} \\
\num{1.85} & \num{2.2} & \num{7.66} & \num{1.47}  \\
\num{2.14} & \num{2.8} & \num{8.41} & \num{1.48}  \\
\br
\end{tabular}
\end{indented}
\end{table}

\begin{figure}[!h] %Modificar
    \centering
    \includegraphics[width=0.57\linewidth]{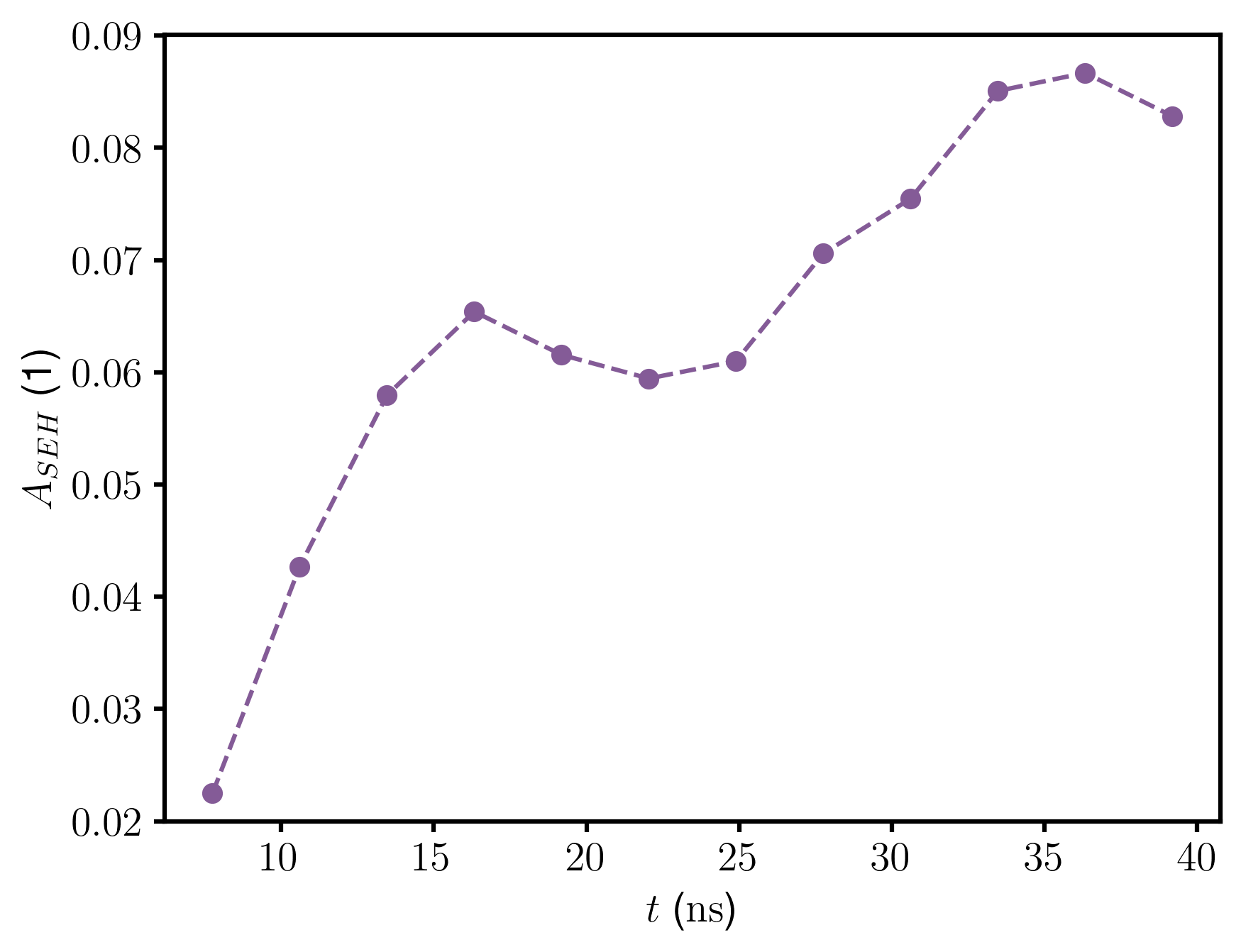} 
    \caption{Evolution of parameter $A_{SEH}$ in time after scenario 3.} %Explicar más condiciones de este resultado
    \label{fig:A_SEH}
\end{figure}

Finally, the temporal evolution of the stochastic electron heating parameter at the location where the upper hybrid frequency is detected is depicted in Figure \ref{fig:A_SEH} for Scenario 3. The considered time interval started at the oscillation period \num{19} because, at this point, radiation absorption (X-B stage) began, which was defined based on the visualization of the radiation electric field in the plasma. The behavior of the $A_{SEH}$ parameter was tracked over \num{96} oscillation periods. Throughout this period, the curve consistently satisfies the condition $|A_{SEH}| < 1$. This result indicates that the electrons in the SCR-1 plasma did not exhibit chaotic motion, even though the amplitude of the electric field of the electrostatic waves may have been sufficiently small to prevent synchronized electron motion \cite{senstius2023}.

\section{Conclusion}\label{sec:conclusion}
The present work examines the inaugural physics inquiries conducted in the SCR-1 stellarator, focusing on fundamental Langmuir probe measurements, MHD equilibrium calculations, and the potential for generating electron Bernstein waves through O-X-B mechanism conversion. This study has yielded results that address the feasibility and inherent challenges of this process.

The SCR-1 stellarator consists of a toroidal vacuum vessel equipped with twenty-six ports and five primary systems, including power, gas injection, ECR heating, and a straightforward Langmuir probe. Successful plasma discharges were achieved during an experimental campaign conducted from March to April 2023.

During the plasma discharge process, the medium was characterized by measuring the electron temperature and density using a Langmuir probe. Adjustments were made to the I-V characteristic curve to account for plasma magnetization. Despite these corrections, electron density values of \SI{1.21e16}{\meter^{-3}} and an electron temperature of \SI{15.1}{\electronvolt} were obtained in the plasma core, which closely matched the theoretical predictions based on the particle and energy balance.

Magnetohydrodynamic equilibrium and linear stability calculations utilizing the VMEC code were employed to characterize the experimental parameters of SCR-1. The findings revealed a low total beta parameter of approximately \SI{0.02}{\percent} when considering medium-pressure, reduced aspect ratio, and magnetic field flux surfaces with a width of approximately \SI{0.12}{\cm}. The SCR-1 stellarator exhibited rational rotational transform values, magnetic islands, and reversed magnetic shear, which significantly enhanced confinement. Furthermore, the linear stability of the SCR-1 stellarator plasma was demonstrated by the presence of a magnetic well and a positive Mercier term over a significant portion of the magnetic flux surfaces.

The plasma characterization process plays a vital role in generating input files for the IPF-FMDC wave code. These files included \num{11} electron density and magnetic field files as well as considerations for incident radiation, such as launch angles, beam width, and oscillation periods. To streamline the simulations, it was necessary to perform multiple parallel tasks on cicima-hpc clusters. Three scenarios were investigated for the O-X-B mechanism, with two of them operating in single-path mode and one including the vacuum vessel geometry.

In scenario 1, the electron density was adjusted to create an overdense plasma while maintaining the same experimental electron temperature. This allowed the identification of an ordinary mode cutoff region, and three stages of the O-X-B mechanism were observed, with three conversion zones being identified. The primary stage of O-X conversion demonstrated a maximum value of \SI{13}{\percent} with an electron density to cutoff electron density ratio of \num{2.14} over a width of $\SI{0.7} \lambda_0$. The rationale behind this value can be attributed to various factors outlined by the theory, including the normalized electron density scale lengths, magnetic field strength, coupling, and mode conversion radius of curvature. These parameters were consistent with previous findings. The presence of plasma curvature has an adverse impact on the O-X conversion percentage, resulting in a decrease in its value. The conversion rate from O to X declines dramatically from \SI{100}{\percent}. The angular window exhibited a more pronounced correlation with a greater magnitude as a result of the elevated density ratio. This effect also eliminates the reliance on the toroidal and poloidal orientations of the incident radiation to determine the optimal angular window for the O-X-B conversion mechanism.

In Scenario 2, similar phenomena were observed when the magnitude of the magnetic field components was altered owing to the characteristic scales. The magnetic field of SCR-1 does not play a significant role in electron Bernstein waves within the selected range.

In Scenario 3, the O-X conversion percentage was examined by considering multiple reflections from the SCR-1 vacuum vessel. This analysis considered the magnitude of the magnetic field utilized in plasma discharges as well as the reflection of radiation in the vacuum vessel. A conversion rate of \SI{63}{\percent} for O-X conversion was achieved at the highest electron density ratio, with a clear dependence on the toroidal angle. A closer alignment was observed between the radiation and plasma curvature, and an additional conversion region was observed near the electronic cyclotron heating region.

The damping mechanism is important for X-B conversion. The conversion of SX-FX had a more pronounced effect, as none of the four conversion regions exhibited a percentage below \SI{40}{\percent}. The observed behavior of the $A_{SEH}$ parameter over \num{96} oscillation periods is that the electrons in the SCR-1 plasma did not exhibit chaotic motion.  Furthermore, it was confirmed that ion-electron and neutral-electron collisions do not lead to damping of electron Bernstein waves.

Finally, the current configuration of the SCR-1 stellarator did not exhibit a plasma region with a cutoff for the ordinary mode, limiting the feasibility of the O-X-B mechanism. Despite this, an increase in electron density, achievable within the limits of its power system, allowed the identification of four zones for O-X conversion using the full-wave code, considering multiple reflections from the vacuum chamber. The fraction converted to the slow extraordinary mode decreases in the upper hybrid region, and the low electric field amplitude might not enable coordinated electron gyration for the propagation of B modes in the plasma.

Future research should explore modifications in the radial profile of the electron density to achieve a normalized electron density scale length greater than one. This can be accomplished by adjusting the delivered power or the confining magnetic field of the plasma, always considering the maximum beta value that the SCR-1 plasma can attain. This adjustment could be implemented using the magnetohydrodynamic equilibrium code presented in this work, along with the characterization of the nonlinear stability of the plasma and the effects of currents in the plasma through the computational suite STELLOPT. Additionally, it is necessary to obtain a power deposition profile in the conversion zones either through a computational code that considers finite Larmor radius effects or experimentally. This will allow for an in-depth study of stochastic electron heating and parametric instabilities, which is another nonlinear phenomenon of waves in the plasma.

Another suggestion is to use a gyrokinetic code to study neoclassical plasma transport. This would provide a better understanding of turbulence calculations in the plasma at Larmor radius scales and the transport coefficients of particles and energy owing to changes in the medium parameters. This enables the construction of new input files for the full-wave code used in this study.

\section*{References}
\bibliographystyle{iopart-num} 
\bibliography{ref}

\end{document}